\documentclass[11pt]{JHEP3}

\usepackage{epsfig}
\usepackage{citesort}

\def  \mlh    {\hat{m}_\ell}
\def  \mlhs   {\hat{m}_\ell^2}
\def  \mkh    {\hat{m}_{K^*}}
\def  \mkhs   {\hat{m}_{K^*}^2}
\def  \sh     {\hat{s}}

\title{\boldmath Lepton polarization asymmetries for $B \to K^*
\ell^+ \ell^-$:\\ 
A Model Independent approach}

\author{A. S. Cornell 
\footnote{Present address:
Yukawa Institute for Theoretical Physics, 
Kyoto University, 
Kitashirakawa Oiwake-Cho, Sakyo-ku, 
Kyoto 606-8502, Japan.} \\
 Korea Institute of Advanced Study, 207-43 Cheongryangri 2-dong,\\
 Dongdaemun-gu, Seoul 130-722, Korea.\\
E-mail : alanc@yukawa.kyoto-u.ac.jp}
\author{Naveen Gaur \\
Department of Physics \& Astrophysics \\
University of Delhi, Delhi - 110 007, India.\\
E-mail : naveen@physics.du.ac.in \\}

\preprint{
{\tt hep-ph/0408164}
}

\abstract{
\par In this work we shall derive expressions for the single and
double lepton polarization asymmetries for the exclusive decay $B
\to K^* \ell^+ \ell^-$, using the most general model independent
effective Hamiltonian. We have conducted this study with this
particular channel as it has the highest branching ratio among the
various purely leptonic and semi-leptonic decay modes, making this
mode particularly useful for studying physics beyond the SM. We
have also analyzed the effects on these polarization asymmetries,
and hence the physics underlying it, when complex phases are
included in some of the Wilson coefficients.
}


\begin{document}

%%%%%%%%%%%%%%%%%%%%%%%
%  Section : 1

\section{Introduction}\label{section:1}

\par As more and more experimental data is produced by
$B$-factories our quest for finding new physics signatures in the
various decay modes for low energy processes is increasing. The
sheer volume of literature studying the possible signatures of
different supersymmetric (and other) models in the context of
$B$-meson decays evidences how promising a testing ground these
rare decays, induced by the flavour changing neutral current
(FCNC) $b \to s$, are. Of the various hadronic, leptonic and
semi-leptonic decays modes (based on the $b \to s$ transition of
the $B$-meson) the semi-leptonic decay modes are extremely
significant as they are theoretically cleaner, and hence very
useful for testing various new physics models. 
The semi-leptonic decay modes based on the quark level transition $b
\to s \ell^+ \ell^-$ offer many more observables associated with the final
state lepton pair, such as the  forward-backward (FB) asymmetry,
lepton polarization asymmetries etc. These additional observables
could prove to be very useful in testing the effective structure of
these theories and hence the underlying physics. For this reason many
processes like $B \to \pi (\rho) \ell^+ \ell^-$ 
\cite{Choudhury:2002fk}, $B \to \ell^- \ell^+ \gamma$
\cite{RaiChoudhury:2002hf}, $B \to K \ell^+ \ell^-$
\cite{Choudhury:2003xg} and the inclusive process $B \to X_s \ell^+ 
\ell^-$ \cite{Kruger:1996cv,Fukae:1998qy,RaiChoudhury:1999qb} have
been studied. But of the various decay modes of the
$B$-mesons based on the transition $b \to s \ell^+ \ell^-$ the
exclusive mode $B \to K^* \ell^+ \ell^-$ is one of the more attractive
due to it having the highest standard model (SM) branching ratio. For
this reason large numbers of observables in this decay mode have been
studied
\cite{Ali:1999mm,Aliev:2000jx,Burdman:1995ks,Aliev:2004hi,Aliev:2002ux}.

\par Previously Aliev {\em et al.} \cite{Aliev:2000jx} studied the
various single polarization asymmetries for this decay mode, where
they used the model independent approach earlier proposed by Fukae
{\em et 
al.} \cite{Fukae:1998qy}. They were able to demonstrate that within
the framework of a model independent theory, constrained by the
experimentally measured values of the $B \to K^* \ell^+ \ell^-$
branching ratio, there existed regions where the possible new Wilson
coefficients could generate considerable departures from the
SM. However, as pointed out in London {\em et al.}
\cite{Bensalem:2002ni} some of the single lepton polarization 
asymmetries may be too small to be observed, and hence merely the
single lepton polarization asymmetries may not provide a
sufficient number of observables to crosscheck the structure of
the effective Hamiltonian. With this in mind more observables are
required.

\par Further to this there has been in the recent observations of
the $B \to \pi \pi$ and $B \to \pi K$ decays hints of possible
anomalies unexplainable within the SM
\cite{Buras:2000gc,Yoshikawa:2003hb}. These  
anomalies arise when we try to match the pattern of data from the
$B$-factories with theory. The recent Belle and BaBar data
regarding the $B \to \pi \pi$ mode can be easily explained by
taking into account the non-factorizable contributions. Note that
the $B \to \pi \pi$ channel is not greatly effected by the
electroweak (EW) penguin diagrams and therefore one can extract
the hadronic parameters from this by assuming isospin symmetry.
Using the SU(3) flavour symmetry we can determine the hadronic $B
\to \pi K$ parameters from the relevant $B \to \pi \pi$ modes. As
has been pointed out some time back by Buras {\sl et al.}
\cite{Buras:2000gc}, which has been revived in many later works
\cite{Yoshikawa:2003hb}, this 
procedure works very well and gives us a good match between theory and
experimental results as long as we are analyzing those modes which
are not greatly affected by the EW penguin diagrams. However, if
we try to repeat the same sort of exercise for modes like $B_d \to
\pi^0 K_S$, which are dominated by EW penguins, then there is a
substantial disagreement between theory and experimental data
\cite{Buras:2000gc}. Lately some solutions of this {\it ``$ B \to \pi
K$ puzzle''} are being tested and almost all of these propose EW 
penguins which are sizably enhanced not only in magnitude but
also in their CP-violating phase, which can become as large as
-90$^o$. This proposal is a very interesting one and
can significantly affect many other decay modes. Rather detailed
studies of this proposal have been carried out by Buras {\em et
al.} \cite{Buras:2000gc} leading to possible predictions of
substantial enhancements in the branching ratio of many leptonic and 
semi-leptonic decay modes, which will soon be tested in
B-factories. This sort of possibility forces us to consider the
option of what could be the possible changes expected in various
kinematical observables, such as the branching ratios, FB
asymmetries and various polarization asymmetries, if some of the
Wilson coefficients had such a large phase (making them
predominately imaginary). Note that with this in mind we have
analyzed this in an earlier work \cite{RaiChoudhury:2004pw} for the
inclusive decay mode $B \to X_s \ell^+ \ell^-$. In that study we
estimated the variation in the polarization asymmetries in the
inclusive mode if the $b s Z$ vertex were modified. In this work we also
explored the option of allowing some of the Wilson coefficients
having large CP violating phases. This sort of approach to finding the
effects of extra phases in Wilsons on various kinematical observables
like branching ratio, partial width  
CP asymmetry, FB asymmetry and single lepton polarization asymmetry in
$B \to K^* \ell^+ \ell^-$ have been followed in earlier works
\cite{Kruger:2000zg}. In this current study we shall also try to
analyze what effects there shall be on the various polarization
asymmetries in the $B \to K^* \ell^+ \ell^-$ decay. 

\par In this study we will work in a model independent framework
by taking the most general form of the effective Hamiltonian and
then analyzing the effect on polarization asymmetries if the
Wilson coefficients (mainly the coefficients which correspond to
vector like interactions) have an extra phase. Keeping this
eventual aim in mind, this paper shall be organized as follows: In
section 2 we shall introduce the most general form of the
effective Hamiltonian, obtaining (in terms of the forms factors
for the $B \to K^*$ transition) the matrix element for the $B \to
K^* \ell^+ \ell^-$ decay and the unpolarized cross-section. In
section 3 we shall define and calculate the various single and
double polarization asymmetries, followed in section 4 with our
numerical analysis. We shall also include a discussion of these
results and our conclusions in this final section.

%%%%%%%%%%%%%%%%%%%%%%%%%%%%%%%%%%
%  Section : 2

\section{The Effective Hamiltonian}\label{section:2}

\par We know, from the paper by Fukae {\em et al.} \cite{Fukae:1998qy}
that together with the terms proportional to our conventionally
defined $C_7$ (written below as $C_{SL}$ and $C_{BR}$ for terms
corresponding to the standard $-2 m_s C_7$ and $-2 m_b C_7$ terms
respectively), $C_9$ and $C_{10}$ (which can be redefined in terms of
$C_{LL}$ and $C_{LR}$) we have ten independent local four-Fermi
interactions which contribute to the FCNC transition $b \to s
\ell^+ \ell^-$;
\begin{eqnarray}
{\cal H}_{eff} & = & \frac{\alpha G_F}{\sqrt{2} \pi} V_{ts}^*
V_{tb} \left[ C_{SL} \left( \bar{s} i \sigma_{\mu\nu}
\frac{q^{\nu}}{q^2} L b \right) \left( \bar{\ell} \gamma^{\mu}
\ell \right) + C_{BR} \left( \bar{s} i \sigma_{\mu\nu}
\frac{q^{\nu}}{q^2} R b \right) \left( \bar{\ell} \gamma^{\mu}
\ell \right) \right. \nonumber \\
&& \hspace{1in} + C_{LL} \left( \bar{s}_L \gamma_{\mu} b_L \right)
\left( \bar{\ell}_L \gamma^{\mu} \ell_L \right) + C_{LR} \left(
\bar{s}_L \gamma_{\mu} b_L \right) \left( \bar{\ell}_R
\gamma^{\mu} \ell_R
\right) \nonumber \\
&& \hspace{1in} + C_{RL} \left( \bar{s}_R \gamma_{\mu} b_R \right)
\left( \bar{\ell}_L \gamma^{\mu} \ell_L \right) + C_{RR} \left(
\bar{s}_R \gamma_{\mu} b_R \right) \left( \bar{\ell}_R
\gamma^{\mu} \ell_R \right) \nonumber \\
&& \hspace{1in} + C_{LRLR} \left( \bar{s}_L b_R \right) \left(
\bar{\ell}_L \ell_R \right) + C_{RLLR} \left( \bar{s}_R b_L
\right) \left( \bar{\ell}_L \ell_R \right) \nonumber \\
&&  \hspace{1in} + C_{LRRL} \left( \bar{s}_L b_R \right) \left(
\bar{\ell}_R \ell_L \right) + C_{RLRL} \left( \bar{s}_R b_L
\right) \left( \bar{\ell}_R \ell_L \right) \nonumber \\
&& \hspace{1in} + C_T \left( \bar{s} \sigma_{\mu\nu} b \right)
\left( \bar{\ell} \sigma^{\mu\nu} \ell \right) + i C_{TE} \left(
\bar{s} \sigma_{\mu\nu} b \right) \left( \bar{\ell}
\sigma_{\alpha\beta} \ell \right) \epsilon^{\mu\nu\alpha\beta}
\Bigg] , \label{effH}
\end{eqnarray}
where $q$ represents the momentum transfer ($q = p_B - p_{K^*}$),
$L/R = (1 \mp \gamma^5 )/2$ and the $C_X$'s are the coefficients
of the four Fermi interactions. Among these there are four vector
type interactions ($C_{LL}$, $C_{LR}$, $C_{RL}$ and $C_{RR}$), two
of which contain contributions from the SM Wilson coefficients.
These two coefficients, $C_{LL}$ and $C_{LR}$, can be written as;
\begin{eqnarray}
C_{LL}^{tot} & = & C_9 - C_{10} + C_{LL} , \nonumber \\
C_{LR}^{tot} & = & C_9 + C_{10} + C_{LR} . \label{C9C10}
\end{eqnarray}
So $C_{LL}^{tot}$ and $C_{LR}^{tot}$ describe the sum of the
contributions from the SM and new physics. 
Eqn.(\ref{effH}) also contains four scalar type interactions
($C_{LRLR}$, $C_{RLLR}$, $C_{LRRL}$ and $C_{RLRL}$) and two tensor
type interactions ($C_T$ and $C_{TE}$).

\par We shall now follow the standard techniques, as seen in
references \cite{Fukae:1998qy,Ali:1999mm,Aliev:2000jx,Burdman:1995ks} 
of rendering the quark level transition
above to a matrix element which describes the exclusive process $B
\to K^* \ell^+ \ell^-$, that is, by parameterizing over the $B$
and $K^*$ meson states in terms of form factors. Using the form
factor expressions derived in the paper by Ball {\em et
al.} \cite{Ball:1998kk} we express our hadronic matrix elements as;
\begin{eqnarray}
\langle K^* | \bar{s} ( 1 \pm \gamma^5 ) b | B \rangle & = & \mp
\frac{ 2 i m_{K^*}}{m_b} ( \epsilon^* \cdot q ) A_0(\hat{s}) ,
\label{form1}
\end{eqnarray}
\begin{eqnarray}
\langle K^* | \bar{s} i \sigma_{\mu\nu} q^{\nu} ( 1 \pm \gamma^5 )
b | B \rangle & = & -2 \epsilon_{\mu\nu\rho\sigma} \epsilon^{*
\nu} q^{\rho} p_{K^*}^{\sigma} T_1 (\hat{s}) \pm i T_2 (\hat{s})
\left[ \epsilon^{*}_{\mu} \left( m_B^2 - m_{K^*}^2 \right) -
(\epsilon^* \cdot q ) \right. \nonumber \\ 
&& \left. \left( 2 p_{K^*} + q \right)_{\mu} \right]
 \pm i T_3 (\hat{s}) \left( \epsilon^* \cdot q
\right) \left[ q_{\mu} - \frac{q^2}{m_B^2 - m_{K^*}^2} \left( 2
p_{K^*} + q \right)_{\mu} \right] , \label{form2}
\end{eqnarray}
\begin{eqnarray}
\langle K^* | \bar{s} \gamma_{\mu} ( 1 \pm \gamma^5 ) b | B
\rangle & = & \epsilon_{\mu\nu\alpha\beta} \epsilon^{* \nu}
q^{\alpha} p_{K^*}^{\beta} \left( \frac{2 V(\hat{s})}{m_B +
m_{K^*}} \right) \pm i \epsilon^{*}_{\mu} \left( m_B + m_{K^*}
\right) A_1(\hat{s}) \nonumber \\
&& \hspace{0.5in} \mp i \left( 2 p_{K^*} + q \right)_{\mu} \left(
\epsilon^* \cdot q \right) \frac{ A_2(\hat{s}) }{ m_B + m_{K^*} }
\nonumber \\
&& \hspace{0.5in} \mp i q_{\mu} \left( \epsilon^* \cdot q \right)
\frac{2 m_{K^*}}{\hat{s}} \left( A_3(\hat{s}) - A_0(\hat{s})
\right) , \label{form3}
\end{eqnarray}
\begin{eqnarray}
\langle K^* | \bar{s} \sigma_{\mu\nu} b | B \rangle & = & - i
\epsilon_{\mu\nu\alpha\beta} \left[ T_1 (\hat{s}) \epsilon^{*
\alpha} \left( 2 p_{K^*} + q \right)^{\beta} - \frac{ \left( m_B^2
- m_{K^*}^2 \right) }{q^2} \left\{ T_1 (\hat{s}) - T_2 (\hat{s})
\right\} \epsilon^{* \alpha} q^{\beta} \right. \nonumber \\
&& \hspace{1in} + \frac{ 2 \left( \epsilon^* \cdot q \right)}{q^2}
\left\{ T_1 (\hat{s}) - T_2 (\hat{s}) - \frac{q^2}{m_B^2 -
m_{K^*}^2} T_3 (\hat{s}) \right\} p_{K^*}^{\alpha} q^{\beta}
\Bigg] . \nonumber \\ \label{form4}
\end{eqnarray}
Note that the parameterization of these form factors can be found
in Appendix \ref{formfactor}.

\par Using these form factor expressions our matrix element for
the decay $B \to K^* \ell^+ \ell^-$ can be expressed as;
\begin{eqnarray}
{\cal M} ( B \to K^* \ell^+ \ell^- ) & = & \frac{\alpha
G_F}{4 \sqrt{2} \pi} V_{tb} V_{ts}^*
\Bigg[ \left(\bar{\ell} \gamma^{\mu} \ell \right)
\left\{ A \epsilon_{\mu\nu\rho\sigma} \epsilon^{* \nu}
q^{\rho} p_{K^*}^{\sigma} +
i B \epsilon^{*}_{\mu }
+ 2 i ~C ~ (p_{K^*})_\mu \left( \epsilon^* . q \right) \right\}
    \nonumber \\
&& \hspace{.2in} + \left( \bar{\ell} \gamma^{\mu} \gamma_5 \ell
\right) \left\{ E ~ \epsilon_{\mu\nu\rho\sigma} \epsilon^{* \nu }
q^{\rho} p_{K^*}^{\sigma} + i ~F \epsilon^{*}_{\mu } + 2 i ~ G ~
p_{K^* \mu} \left( \epsilon^* . q \right)
\right\} \nonumber \\
&& \hspace{.2in} +  i ~ K ~ \left(\bar{\ell} ~ \ell \right) \left(
\epsilon^* . q \right) + i ~ M ~ \left(\bar{\ell} \gamma_5 \ell
\right)
\left( \epsilon^* . q \right) \nonumber \\
&& \hspace{.2in} + 4 i C_T \left( \bar{\ell} \sigma^{\mu\nu} \ell
\right) \epsilon_{\mu\nu\rho\sigma} \Bigg\{ - 2 T_1 \epsilon^{*
\rho } p_{K^*}^\sigma + N_1 \epsilon^{* \rho } q^\sigma - N_2
(\epsilon . q) p_{K^*}^\rho q^\sigma \Bigg\}
\nonumber \\
&& \hspace{.2in} + 16  C_{TE} \left( \bar{\ell} \sigma^{\mu\nu}
\ell \right) \Bigg\{ - 2 T_1 \epsilon^{* \mu} p_{K^*}^\nu + M_1
\epsilon^{* \mu } q^\nu - M_2 (\epsilon . q) p_{K^*}^\mu q^\nu
\Bigg\} \Bigg] , \nonumber \\ \label{matrixelem}
\end{eqnarray}
where;
\begin{eqnarray}
A & = & \left( C_{LL}^{tot} + C_{RL} \right) \frac{V(\hat{s})}{m_B +
m_{K^*}} - 4 \left( C_{SL} + C_{BR} \right) \frac{T_1(\hat{s})}{q^2} ,
\nonumber \\
B & = & \left( C_{RL} + C_{RR} - C_{LL}^{tot} - C_{LR}^{tot} \right)
\left(m_B + m_{K^*} \right) A_1
- 2 \left( C_{BR} - C_{SL} \right)
\frac{T_2(\hat{s})}{q^2} \left( m_B^2 - m_{K^*}^2 \right)
, \nonumber \\
C & = & \left(C_{LL}^{tot} + C_{LR}^{tot} - C_{RL} - C_{RR} \right)
\frac{A_2(\hat{s})}{m_B + m_{K^*}}
- 2 \left(C_{BR} - C_{SL} \right) \frac{1}{q^2}
\Bigg[T_2(\hat{s}) + \frac{q^2}{(m_B^2 - m_{K^*}^2)} T_3(\hat{s})
\Bigg] , \nonumber \\
D & = & 2 \left(C_{LL}^{tot} + C_{LR}^{tot} - C_{RL} - C_{RR}
\right) \frac{m_{K^*}}{q^2} \left( A_3 (\hat{s}) - A_0(\hat{s})
\right) + 2 \left( C_{BR} - C_{SL} \right)
\frac{T_3(\hat{s})}{q^2} ,
      \nonumber \\
&& + \left(C_{LL}^{tot} + C_{LR}^{tot} - C_{RL} - C_{RR} \right)
+ \frac{A_2(\hat{s})}{m_B + m_{K^*}}
- 2 \left(C_{BR} - C_{SL} \right) \frac{1}{q^2}
\Bigg[T_2(\hat{s}) + \frac{q^2}{(m_B^2 - m_{K^*}^2)} T_3(\hat{s})
\Bigg]   \nonumber \\
E & = & 2 \left( C_{RR} + C_{LR}^{tot} -C_{LL}^{tot} - C_{RL} \right)
\frac{V(\hat{s})}{(m_B + m_{K^*})}
, \nonumber \\
F & = & \left( C_{RR} - C_{LR}^{tot} \right) \left( m_B
+ m_{K^*} \right) A_1(\hat{s})  , \nonumber \\
G & = &  \left( C_{RL} + C_{LR}^{tot} - C_{LL}^{tot} - C_{RR} \right)
\frac{A_2(\hat{s})}{(m_B + m_{K^*})} , \nonumber \\
H & = & 2 \left( C_{LR}^{tot} + C_{RL} - C_{LL}^{tot} - C_{RR} \right)
\frac{m_{K^*}}{q^2} \left( A_3(\hat{s}) - A_0(\hat{s}) \right)
+ \left( C_{RL} + C_{LR}^{tot} - C_{LL}^{tot} - C_{RR} \right)
\frac{A_2(\hat{s})}{(m_B + m_{K^*})}, \nonumber \\
K &=& 2  \left( C_{RLLR} + C_{RLRL} - C_{LRLR} - C_{LRRL} \right)
\frac{m_{K^*}}{m_b} A_0(\hat{s})
, \nonumber \\
M &=& 2  \left( C_{LRRL} + C_{RLLR} - C_{LRLR} - C_{RLRL}\right)
\frac{m_{K^*}}{m_b} A_0(\hat{s})  , \nonumber \\
N_1 & = & - T_1(\hat{s}) + \frac{\left( m_B^2 -
m_{K^*}^2 \right)}{q^2} \left\{ T_1(\hat{s}) - T_2(\hat{s})
\right\} , \nonumber \\
N_2 & = & \frac{2}{q^2} \left[ T_1(\hat{s}) - T_2(\hat{s})
- \frac{q^2}{(m_B^2 - m_{K^*}^2)} T_3(\hat{s}) \right] , \nonumber \\
M_1 &=&  N_1 , \nonumber \\
M_2 &=&  N_2 .
\label{consts}
\end{eqnarray}
\par Using the above expression we can calculate the unpolarized
decay rate as;
\begin{eqnarray}
\frac{d\,\Gamma}{d\,\hat{s}} \left( B \to K^* \ell^+ \ell^-
\right) & = & \frac{G_F^2 \alpha^2 }{2^{14} \pi^5 m_B}
\left| V_{ts} V_{tb}^* \right|^2
\lambda^{1/2} \sqrt{1 - \frac{4 m_\ell^2}{q^2}}
~ \Delta ,
\label{sec:2:eqn:9}
\end{eqnarray}
where $\lambda = 1 + \hat{m}_{K^*}^4 + \hat{s}^2 - 2
(\hat{m}_{K^*} + \hat{s} ) - 2 \hat{m}_{K^*} \hat{s} $ with
$\hat{m}_{K^*} = m_{K^*}/m_B$ and $\hat{s} = s/m_B^2$. $s$ is the
dilepton invariant mass. The function $\Delta$ is defined in
Appendix \ref{delta}.

%%%%%%%%%%%%%%%%%%%%%%%%%%%%%%%%%%
%  Section : 3

\section{Lepton polarization asymmetries}\label{section:3}

\par In order to now calculate the polarization asymmetries of
both the leptons defined in the effective four fermion interaction
of Eqn.(\ref{effH}), we must first define the orthogonal vectors
$S$ in the rest frame of $\ell^-$ and $W$ in the rest frame of
$\ell^+$ (where these vectors are the polarization vectors of the
leptons). Note that we shall use the subscripts $L$, $N$ and $T$
to correspond to the leptons being polarized along the
longitudinal, normal and transverse directions respectively
\cite{Kruger:1996cv,Aliev:2000jx,Bensalem:2002ni,RaiChoudhury:1999qb,RaiChoudhury:2002hf}.
\begin{eqnarray}
S^\mu_L & \equiv & (0, \mathbf{e}_{L}) ~=~ \left(0,
\frac{\mathbf{p}_-}{|\mathbf{p}_-|}
\right) , \nonumber \\
S^\mu_N & \equiv & (0, \mathbf{e}_{N}) ~=~ \left(0,
\frac{\mathbf{p_{K^*}} \times \mathbf{p}_-}{|\mathbf{p_{K^*}} \times
\mathbf{p}_- |}\right) , \nonumber \\
S^\mu_T & \equiv & (0, \mathbf{e}_{T}) ~=~ \left(0, \mathbf{e}_{N}
\times \mathbf{e}_{L}\right) , \label{sec3:eq:1} \\
W^\mu_L & \equiv & (0, \mathbf{w}_{L}) ~=~ \left(0,
\frac{\mathbf{p}_+}{|\mathbf{p}_+|} \right) , \nonumber \\
W^\mu_N & \equiv & (0, \mathbf{w}_{N}) ~=~ \left(0,
\frac{\mathbf{p_{K^*}} \times \mathbf{p}_+}{|\mathbf{p_{K^*}} \times
\mathbf{p}_+ |} \right) , \nonumber \\
W^\mu_T & \equiv & (0, \mathbf{w}_{T}) ~=~ (0, \mathbf{w}_{N}
\times \mathbf{w}_{L}) , \label{sec3:eq:2}
\end{eqnarray}
where $\mathbf{p}_+$, $\mathbf{p}_-$ and $\mathbf{p_{K^*}}$ are
the three momenta of the $\ell^+$, $\ell^-$ and $K^*$ particles
respectively. On boosting the vectors defined by
Eqns.(\ref{sec3:eq:1},\ref{sec3:eq:2}) to the c.m. frame of the
$\ell^- \ell^+$ system only the longitudinal vector will be
boosted, whilst the other two vectors remain unchanged. The
longitudinal vectors after the boost will become;
\begin{eqnarray}
S^\mu_L & = & \left( \frac{|\mathbf{p}_-|}{m_\ell}, \frac{E_{\ell}
\mathbf{p}_-}{m_\ell |\mathbf{p}_-|} \right) , \nonumber \\
W^\mu_L & = & \left( \frac{|\mathbf{p}_-|}{m_\ell}, -
\frac{E_{\ell} \mathbf{p}_-}{m_\ell |\mathbf{p}_-|} \right) .
\label{sec3:eq:3}
\end{eqnarray}
The polarization asymmetries can now be calculated using the spin
projector ${1 \over 2}(1 + \gamma_5 \!\!\not\!\! S)$ for $\ell^-$
and the spin projector ${1 \over 2}(1 + \gamma_5\! \not\!\! W)$
for $\ell^+$.

\par Equipped with the above expressions we now define the various
single lepton and double lepton polarization asymmetries. Firstly,
the single lepton polarization asymmetries are defined as
\cite{Kruger:1996cv,Aliev:2000jx,Bensalem:2002ni,RaiChoudhury:1999qb,RaiChoudhury:2002hf};
\begin{eqnarray}
{\cal P}_x^- & \equiv & \frac{\left(\frac{d\Gamma( S_x, W_x
)}{d\hat{s}} + \frac{d\Gamma( S_x, - W_x )}{d\hat{s}} \right) -
\left( \frac{d\Gamma( - S_x, W_x )}{d\hat{s}} + \frac{d\Gamma( -
S_x, - W_x )}{d\hat{s}} \right)} {\left( \frac{d\Gamma( S_x, W_x
)}{d\hat{s}} + \frac{d\Gamma( S_x, - W_x )}{d\hat{s}} \right) +
\left( \frac{d\Gamma( - S_x, W_x )}{d\hat{s}} + \frac{d\Gamma( -
S_x, - W_x )}{d\hat{s}} \right)}, \nonumber \\
{\cal P}_x^+ & \equiv & \frac{\left( \frac{d\Gamma( S_x, W_x
)}{d\hat{s}} + \frac{d\Gamma( - S_x, W_x )}{d\hat{s}} \right) -
\left( \frac{d\Gamma( S_x, - W_x )}{d\hat{s}} + \frac{d\Gamma( -
S_x, - W_x )}{d\hat{s}} \right)} {\left( \frac{d\Gamma( S_x, W_x
)}{d\hat{s}} + \frac{d\Gamma( S_x, - W_x )}{d\hat{s}} \right) +
\left( \frac{d\Gamma( - S_x, W_x )}{d\hat{s}} + \frac{d\Gamma( -
S_x, - W_x )}{d\hat{s}} \right)} , \label{sec3:eq:4}
\end{eqnarray}
where the sub-index $x$ can be either $L$, $N$ or $T$. ${\cal
P}^\pm$ denotes the polarization asymmetry of the charged lepton
$\ell^\pm$. Along the same lines we can also define the double
spin polarization asymmetries as \cite{Bensalem:2002ni};
\begin{eqnarray}
{\cal P}_{xy} & \equiv & \frac{\left( \frac{d\Gamma( S_x, W_y
)}{d\hat{s}} - \frac{d\Gamma( - S_x, W_y )}{d\hat{s}} \right) -
\left( \frac{d\Gamma( S_x, - W_y )}{d\hat{s}} - \frac{d\Gamma(-
S_x, - W_y )}{d\hat{s}}\right)} {\left( \frac{d\Gamma( S_x, W_y
)}{d\hat{s}} + \frac{d\Gamma( - S_x, W_y )}{d\hat{s}} \right) +
\left( \frac{d\Gamma( S_x, - W_y )}{d\hat{s}} + \frac{d\Gamma( -
S_x, - W_y )}{d\hat{s}}\right)} , \label{sec3:eq:5}
\end{eqnarray}
where the sub-indices $x$ and $y$ can be either $L$, $N$ or $T$.

\par The single lepton polarization asymmetries are then;
\begin{eqnarray}
{\cal P}_L^{(\mp)} &=&
\frac{m_B^2}{\Delta}\sqrt{1 - \frac{4 \mlhs}{\sh}}
\Bigg[
\pm {8 \over 3} m_B^4 \sh \lambda Re(A^* E)
\mp \frac{4}{3 \mkhs}
\left\{\lambda (1 - \mkhs - \sh) Re(B^* G)
\right.         \nonumber \\
&& \left. - \left(\lambda + 12 \sh \mkhs \right) Re(B^* F)
\right\}
\pm \frac{4}{3 \mkhs} m_B^2 \lambda
  \left\{ m_B^2 \lambda Re(C^* G) + (1 - \mkhs - \sh) Re(F^* C)
  \right\}  \nonumber \\
&& + \frac{16}{3 \mkhs} m_B \mlh Re(B^* C_T)
 \left\{ 2 \left(\lambda + 12 \sh \mkhs \right) N_1
      + (\mkhs + \sh - 1)
        \left( m_B^2 \lambda N_2 + 24 \mkhs T_1 \right)
 \right\}     \nonumber \\
&& - \frac{16 \mlh}{3 \mkhs} m_B^3 Re(C^* C_T) \lambda
  \left\{ m_B^2 \lambda N_2 + 2  (\mkhs + \sh - 1) N_1
          + 8 \mkhs T_1
  \right\}    \nonumber \\
&& - \frac{256 \mlh}{3} m_B^3 \lambda T_1
  \left\{Re(A^* C_{TE}) \mp Re(E^* C_T) \right\}
- \frac{4 \mlh}{\mkhs} m_B \lambda
  \left\{ \sh Re(H^* K) + (1 - \sh - \mkhs)
  \right.        \nonumber \\
&& \left. \times  m_B^2 Re(G^* K) + Re(F^* K)
   \right\}
\mp \frac{64 \mlh}{3 \mkhs} m_B \lambda
  \left\{ 2 (\lambda + 12 \sh \mkhs) N_1
      + (\mkhs + \sh - 1)
  \right.   \nonumber \\
&& \left. \times \left( \lambda m_B^2 N_2 + 24 \mkhs T_1 \right)
  \right\} Re(F^* C_{TE})
 \pm \frac{64 \mlh}{3 \mkhs } \lambda m_B^3
   \left\{ \lambda m_B^2 N_2 + 2 (\mkhs + \sh - 1) N_1
   \right.   \nonumber \\
&& \left. + 8 \mkhs T_1
   \right\} Re(G^* C_{TE})
 - 2 \frac{\sh \lambda}{\mkhs} m_B^2 Re(M^* K)
 - \frac{64}{3 \mkhs} m_B^2
  \left\{ m_B^2 \sh \lambda^2 m_B^2 N_2^2
  \right.   \nonumber \\
&& \left.  + 16 m_B^2 \mkhs \sh \lambda
   T_1 N_2 + 64 \mkhs \left(\lambda + 3 \sh \mkhs\right) T_1^2
  \right\}
\Bigg] ,
\label{single-pol:1}  \\
{\cal P}_N^{(\mp)} &=& \frac{\pi m_B^3 \sqrt{\sh \lambda}}{\Delta}
\sqrt{1 - \frac{4 \mlhs}{\sh}}
\Bigg[\pm 2 m_B \mlh \left\{ Im(A^* F) + Im(B^* E) \right\}
- \frac{1}{2 \mkhs}
  \left\{ m_B^2 \lambda \left( Im(K^* C) + Im(M^* G) \right)
  \right.              \nonumber \\
&&\left. + (1 - \sh - \mkhs) \left( Im(K^* B) + Im(K^* F) \right)
\right\}
 + 8 m_B^2 \pi \left\{ \sh N_1 + ( \mkhs + \sh - 1 ) T_1 \right\}
  Im(A^* C_T)   \nonumber \\
&& + 16 T_1 \left\{ 2 Im(C_{TE}^* B) \pm Im(C_T^* F) \right\}
- \frac{\pi \mlh}{\mkhs} \left\{ (\mkhs + \sh - 1) Im(F^* H)
  - 4 \mkhs Im(F^* G) \right\}  \nonumber \\
&& + m_B^3 \frac{\mlh}{\mkhs} \lambda Im(G^* H)
\pm \frac{8 m_B \mlh}{\mkhs}
 \left\{ \lambda m_B^2 N_2 + 2 \left(\mkhs + \sh - 1\right) N_1
  + 8 \mkhs T_1 \right\} Im(C_{TE}^* K)  \nonumber \\
&& \mp 16 m_B^2 \left\{ N_1 \sh + \left( \mkhs + \sh - 1 \right)
  T_1 \right\}
\Bigg] ,
\label{single-pol:2}  \\
{\cal P}_T^{(\mp)} &=& \frac{\pi m_B^2 \sqrt{\sh \lambda}}{\Delta}
\Bigg[
- 4 m_B^2 \mlh Re(A^* B)
\mp  \frac{\mlh \left(\mkhs + \sh - 1\right)}{\mkhs \sh}
\left\{ m_B^2 \sh \left(Re(B^* H) - \lambda Re(C^* H) \right)
\right. \nonumber \\
&& \left. + (1 - \sh - \mkhs) m_B^2 \left( Re(G^* B) - \lambda Re(C^*
G)\right) + \left( Re(F^* B) - \lambda Re(F^* C) \right)
\right\}
- \frac{16 \left(4 \mlhs + \sh \right) \lambda}{\sh}  \nonumber \\
&& \times \left\{ T_1 Re(B^* C_T) -
\left(\sh N_1 + (\mkhs + \sh - 1) T_1
\right) Re(A^* C_{TE})
\right\}
+ \frac{ \left( \sh - 4 \mlhs \right)}{2 \mkhs \sh}
  \left\{ m_B^2 \lambda Re(K^* G) \right.   \nonumber \\
&& \left. + (1 - \sh - \mkhs) Re(F^* K)
  \right\}
 \pm \frac{8 m_B^2}{\sh} \left( \sh N_1 + (\mkhs + \sh - 1) T_1 \right)
 \left\{ m_B^3 \left(\sh - 4 \mlhs\right) Re(C_T^* E)
 \right.    \nonumber \\
&& \left. \mp 128 m_B^2 \mlh T_1 Re(C^*_T C_{TE})
 \right\}
\pm \frac{16 m_B}{\sh \mkhs}
 \left\{ 2 \left(8 \mlhs - \sh \right) \mkhs T_1 + 2 \mlhs \left(\mkhs
+ \sh - 1 \right) N_1 \right.   \nonumber \\
&& \left. + m_B^2 \mlhs \lambda N_2 \right\} Re(C^*_{TE} G)
\mp \frac{16 m_B^3 \mlhs}{\sh \mkhs}
 \left\{ m_B^2 \lambda N_2 + 2 \left( \mkhs + \sh - 1 \right) N_1
 + 8 \mkhs T_1 \right\} Re(M C^*_{TE})
\Bigg] . \nonumber
\\ \label{single-pol:3}
\end{eqnarray}

And the double polarization asymmetries are;
\begin{eqnarray}
{\cal P}_{LL} &=&
 \frac{1}{\Delta} \frac{4 m_B^2}{3 \sh \mkhs}
\Bigg[ \left(2 \mlhs - \sh \right)
\left\{ m_B^4 \sh \mkhs \lambda |A|^2
+ {1 \over 2} \left(\lambda + 12 \sh \mkhs \right) |B|^2
+ \frac{m_B^4}{2} \lambda^2 |C|^2
+ m_B^2
\right.  \nonumber \\
&& \left. \times  \left(1 - \sh - \mkhs\right) \lambda Re(B^* C)
\right\}
- 32 m_B^3 \sh \mkhs \mlh \lambda T_1 Re(C^*_T A)
+ 8 m_B \sh \mlh
\left\{ 2 \left(\lambda + 12 \sh \mkhs\right) N_1
\right.   \nonumber \\
&& \left. + \left(\mkhs + \sh - 1\right) \left(m_B^2 \lambda N_2
+24 \mkhs T_1 \right) Re(C^*_{TE} B)
\right\}
- 8 m_B^3 \sh \mlh \lambda
\left\{ m_B^2 \lambda N_2  + 2 \left(\mkhs + \sh - 1\right) N_1
\right.  \nonumber \\
&& \left. + 8 \mkhs T_1
\right\} Re(C^*_{TE} C)
+ \left(4 \mlhs - \sh\right) m_B^2 \sh \mkhs \lambda
\left\{ m_B^2 |E|^2 + \frac{|K|^2}{4} \right\}
- {1 \over 2}
\left\{ \left(\lambda + 12 \sh \mkhs\right)
\right.  \nonumber \\
&&\left.   - 2 \mlhs \left( 5 \lambda + 24 \sh \mkhs \right)
\right\} |F|^2
- \frac{m_B^4}{2} \lambda
\left\{ \sh \lambda - 2 \mlhs \left( 5 \lambda + 12 \sh \mkhs \right)
\right\} |G|^2
+ m_B^2 \lambda
\left\{ \sh \left(\mkhs + \sh - 1 \right)
\right. \nonumber \\
&& \left. \times Re(F^* G) - 2 \mlhs
  \left( 5 \left(\mkhs + \sh - 1\right) Re(G^* F) - 3 \sh Re(H^* F)
  \right)
\right\}
+ 3 m_B^2 \mlh \lambda \sh
\left\{ \sh Re(M^* H) \right.  \nonumber \\
&&\left. + m_B^2 \left(1 - \sh - \mkhs \right) Re(M^* G)
  + Re(F^* M)
\right\}
+ {3 \over 4} m_B^2 \sh^2 \lambda |M|^2   \nonumber \\
&& + 4 m_B^2 \sh
 \left(\sh - 4 \mlhs\right)
 \left\{ m_B^4 \lambda^2 N_2^2 + 4
  \left(\lambda + 12 \sh \mkhs\right)  N_1^2
 + 4 \left(\mkhs + \sh - 1\right) \right. \nonumber \\
&& \left.
\times \left( m_B^2 \lambda N_1 N_2 + 24 \sh \mkhs N_1 T_1 \right)
 + 16 m_B^6 \sh \mkhs \lambda N_2 T_1
 \right\} |C_T|^2
+ 256 m_B^6 \mkhs
\left\{ \sh \lambda - 6 \mlhs \right. \nonumber \\
&& \left. \times \left(\lambda + 2 \sh \mkhs\right)
\right\} T_1^2 |C_T|^2
 + 4 m_B^2 \left(\sh - 8 \mlhs\right)
 \left\{ m_B^4 \lambda^2 N_2^2 + 4
  \left(\lambda + 12 \sh \mkhs\right) N_1^2
 + 4 \left(\mkhs + \sh - 1\right) \right. \nonumber \\
&& \left.
\times \left( m_B^2 \lambda N_1 N_2 + 24 \sh \mkhs N_1 T_1 \right)
 + 16 m_B^6 \sh \mkhs \lambda N_2 T_1
 \right\} |C_{TE}|^2
+ 512 m_B^6 \mkhs
\left\{ \sh \lambda - 6 \mlhs \right. \nonumber \\
&& \left. \times \left(\mkhs + \sh - 1 \right) \right\} T_1^2
|C_{TE}|^2 \Bigg] ,
\label{pol-double:ll}  \\
%%%%%%%%%%%%%%
{\cal P}_{LN} &=&
\frac{1}{\Delta} \frac{ m_B^2 \pi}{\mkhs} \sqrt{\frac{\lambda}{\sh}}
\Bigg[
\left( \mkhs + \sh - 1 \right) \mlh
\left\{ m_B^2 \left( \sh Im(H^* B) + \left(1 - \sh - \mkhs \right)
Im(G^* B) \right) + Im(F^* B)
\right\}       \nonumber \\
&& + \lambda \mlh m_B^2
\left\{ m_B^2 \left( \sh Im(C^* H) + \left(1 - \sh - \mkhs \right)
Im(C^* G) \right) + Im(C^* F)
\right\} + \frac{m_B \sh}{2}
\left\{ m_B^2 \lambda Im(C^* M) \right.   \nonumber \\
&& \left. + \left(1 - \sh - \mkhs \right) Im(B^* M) \right\}
+ \left(\sh - 4 \mlhs\right)
\left\{ 16 m_B T_1 Im(B^* C_T) + 16 m_B^3
\left( \sh N_1 + \left( \mkhs + \sh - 1\right) T_1 \right)
\right.   \nonumber \\
&& \left. \times Im(C^*_{TE} A)
+ \frac{m_B}{2} \left( \left(\mkhs + \sh - 1\right) Im(F^* K) - m_B^2
\lambda Im(G^* K) \right)
+ 8 m_B^3  \right. \nonumber \\
&& \left. \times \left( \sh N_1 + \left(\mkhs + \sh - 1\right) T_1
\right) Im(E^* C_T)
\right\}
+ 16 m_B
\left\{ \left( 2 \sh \mkhs T_1 + 2 \mlhs \left(\mkhs + \sh - 1\right)
N_1 \right. \right. \nonumber \\
&& \left. \left. + m_B^2 \mlhs \lambda N_2 \right) Im(C^*_{TE} F)
+ m_B^2 \mlhs
  \left( m_B^2 \lambda N_2 + 2 \left( \mkhs + \sh - 1 \right) N_1
   + 8 \mkhs T_1 \right) \right. \nonumber \\
&& \left. \left( \left(\mkhs + \sh - 1\right) Im(G^* C_{TE})
  + \sh Im(C^*_{TE} H) \right)
\right\}
+ 8 m_B^2 \mlh \sh \left\{ m_B^2 \lambda N_2 + 2 \left(\mkhs + \sh - 1
 \right) N_1 \right. \nonumber \\
&& \left. + 8 \mkhs T_1 \right\} \Bigg] ,
\label{pol-double:ln}   \\
%%%%%%%%%%%%
{\cal P}_{LT} &=& \frac{1}{\Delta}
\frac{m_B^2 \pi}{\mkhs}\sqrt{\frac{\lambda}{\sh}} \sqrt{1 - \frac{4
\mlhs}{\sh}}
\Bigg[
- 2 m_B^2 \mkhs \mlh \sh Re(A^* F + B^* E + 8 T_1 C_T^* F)
+ \frac{m_B \sh}{2} \left\{ m_B^2 \lambda
\left(Re(K^* C) \right. \right. \nonumber \\
&& \left. \left. - Re(M^* G) \right) + \left(1 - \sh - \mkhs \right)
\left(Re(B^* K) - Re(M^* F) \right) \right\}
- 8 m_B^3 \mkhs \sh \left\{ \sh N_1 + \left(\mkhs + \sh - 1\right) T_1
\right\}   \nonumber \\
&& \times \left( Re(A^* C_T) - 2 Re(C_T^* E) \right)
 + 32 m_B \sh T_1 Re(B^* C_{TE})
+ \mlhs \sh
\left\{ \left(\mkhs + \sh - 1\right) \left(|F|^2 + m_B^4  \lambda |G|^2
\right)
\right. \nonumber \\
&& \left. - 2 m_B^2 \lambda Re(F^* G)
+ m_B^2 \left(\mkhs + \sh - 1\right) \sh Re(F^* H)
- m_B^4 \sh \lambda Re(G^* H)
\right\}
+ 8 m_B^2 \mlh \sh \left\{ m_B^2 \lambda N_2 \right. \nonumber \\
&& \left. + 2 \left(\mkhs + \sh - 1
\right) N_1 + 8 \mkhs T_1 \right\} Re(K^* C_{TE})
- 256 m_B^2 \mlh \left\{ \sh N_1 + \left(\mkhs + \sh - 1\right) T_1
\right\}    \nonumber \\
&& \left( |C_T|^2 + 4 |C_{TE}|^2 \right) + 16 m_B^3 \sh \left\{\sh
N_1 + \left(\mkhs + \sh - 1\right) T_1 \right\} Re(C_{TE}^* E)
\Bigg] ,
\label{pol-double:lt}   \\
%%%%%%%%%%%%%%%%%%
{\cal P}_{NL} &=&
\frac{1}{\Delta} \frac{ m_B^2 \pi}{\mkhs} \sqrt{\frac{\lambda}{\sh}}
\Bigg[
- \left( \mkhs + \sh - 1 \right) \mlh
\left\{ m_B^2 \left( \sh Im(H^* B) + \left(1 - \sh - \mkhs \right)
Im(G^* B) \right) + Im(F^* B)
\right\}       \nonumber \\
&& - \lambda \mlh m_B^2
\left\{ m_B^2 \left( \sh Im(C^* H) + \left(1 - \sh - \mkhs \right)
Im(C^* G) \right) + Im(C^* F)
\right\} - \frac{m_B \sh}{2}
\left\{ m_B^2 \lambda Im(C^* M) \right.   \nonumber \\
&& \left. + \left(1 - \sh - \mkhs \right) Im(B^* M) \right\}
+ \left(\sh - 4 \mlhs\right)
\left\{ 16 m_B T_1 Im(B^* C_T) + 16 m_B^3
\left( \sh N_1 + \left( \mkhs + \sh - 1\right) T_1 \right)
\right.   \nonumber \\
&& \left. \times Im(C^*_{TE} A)
+ \frac{m_B}{2} \left( \left(\mkhs + \sh - 1\right) Im(F^* K) - m_B^2
\lambda Im(G^* K) \right)
- 8 m_B^3  \right. \nonumber \\
&& \left. \times \left( \sh N_1 + \left(\mkhs + \sh - 1\right) T_1
\right) Im(E^* C_T)
\right\}
- 16 m_B
\left\{ \left( 2 \sh \mkhs T_1 + 2 \mlhs \left(\mkhs + \sh - 1\right)
N_1 \right. \right. \nonumber \\
&& \left. \left. + m_B^2 \mlhs \lambda N_2 \right) Im(C^*_{TE} F)
+ m_B^2 \mlhs
  \left( m_B^2 \lambda N_2 + 2 \left( \mkhs + \sh - 1 \right) N_1
   + 8 \mkhs T_1 \right) \right. \nonumber \\
&& \left. \left( \left(\mkhs + \sh - 1\right) Im(G^* C_{TE})
  + \sh Im(C^*_{TE} H) \right)
\right\}
+ 8 m_B^2 \mlh \sh \left\{ m_B^2 \lambda N_2 + 2 \left(\mkhs + \sh - 1
 \right) N_1 \right. \nonumber \\
&& \left. + 8 \mkhs T_1 \right\} Im(C^*_{TE} M) \Bigg] ,
\label{pol-double:nl}   \\
%%%%%%%%%%%%%%%
{\cal P}_{NN} &=& \frac{1}{\Delta} \frac{2 m_B^2}{3 \sh \mkhs}
\Bigg[ m_B^4 \left(\sh - 4 \mlhs\right) \mkhs \lambda \left( \sh |A|^2
+ \sh |E|^2 + {1 \over 2} |K|^2 \right)
- \left\{ 2 \left(\lambda + 12 \sh \mkhs\right) \mkhs + \sh \lambda
\right\} \left( |B|^2 \right. \nonumber \\
&& \left. - m_B^4 \lambda |G|^2 \right)
+ m_B^2 \lambda \left(\sh + 2 \mlhs\right)
\left\{ m_B^2 \lambda |C|^2 + 2 \left(\mkhs + \sh -1\right)
\left(Re(B^* C) - Re(F^* G)\right) + |F|^2  \right\} \nonumber \\
&& - 48 m_B \mlh \sh \left\{ m_B^2 \lambda N_2  + 2 \left(\mkhs + \sh
- 1\right)  N_1 + 8 \mkhs T_1 \right\}
 \left\{ m_B^2 \lambda Re(C^* C_{TE}) - \left(\mkhs + \sh - 1\right)
\right. \nonumber \\
&& \left. \times  Re\left(B^* C_{TE}\right)  \right\}
+ 2 m_B^2 \sh \mlhs \lambda \left( m_B^2 \sh |H|^2 + 2 Re(F^* H)
+ {1 \over 4} \frac{m_B^2}{\mlhs} |M|^2 \right)
+ 2 m_B^2 \lambda \left(1 - \sh - \mkhs\right) \nonumber \\
&& \times \left\{ \left(2 \mlhs + \sh\right) Re(F^* G) + 6 \sh \mlhs
Re(G^* H)  \right\}  - 6 m_B \mlh \sh \lambda
\left\{ m_B^2 \left( \left(\mkhs + \sh - 1\right) Re(G^* M) \
\right. \right. \nonumber \\
&& \left. \left. - \sh
Re(H^* M) \right) - Re(M^* F) \right\} + 8 m_B^2 \sh \left(\sh - 4
\mlhs\right) |C_T|^2
\left\{ m_B^4 \lambda^2 N_2^2 + 16 m_B^2 \mkhs \lambda T_1 N_2
\right. \nonumber \\
&& \left. + 192
\mkh^4 T_1^2  + 4 \left(\lambda + 12 \sh \mkhs\right) N_1^2
 + 4 \left(\mkhs + \sh - 1\right) \left(m_B^2 \lambda N_2 +
24 \mkhs T_1\right) \right\}  \nonumber \\
&& + 32 m_B^2 \sh |C_{TE}|^2
\left\{ \lambda m_B^2 \left(\sh + 8 \mlhs\right)
\left(\lambda m_B^2 N_2^2
+ 16 T_1 \mkhs T_1 N_2  \right) + 19 \sh \mkh^4 T_1^2
+ 4 \left( 8 \lambda \mlhs \right. \right. \nonumber \\
&& \left. \left. + \sh \left(\lambda + 12 \sh \mkhs \right)
\right) N_1^2 + 4 \left(\mkhs + \sh - 1\right) \left( m_B^2
\lambda \left(\sh + 8 \mlhs\right) N_2 + 24 \sh \mkhs T_1\right)
\right\} \Bigg] ,
\label{pol-double:nn} \\
%%%%%%%%%%%%%%%%%%%%%%%%%%%%%%%%
{\cal P}_{NT} &=& \frac{4}{\Delta} \frac{m_B^2}{3 \mkhs} \sqrt{1 -
\frac{4 \mlhs}{\sh}}
\Bigg[ m_B^4 \sh \lambda Im(E^* A)
+ 32 m_B^3 \mlh \lambda \mkhs T_1
\left( 2 Im(A^* C_{TE}) + Im(E^* C_T) \right)  \nonumber \\
&& + \lambda
\left\{ \left(\mkhs + \sh - 1\right) \left( m_B^2 Im(G^* B) - Im(C^*
F) \right) - Im(F^* B) + m_B^2 \lambda Im(C^* G)
\right\} \nonumber \\
&& + 4 m_B \mlh
\left\{ 2 \left(\lambda + 12 \sh \mkhs\right) N_1
+ \left(\mkhs + \sh - 1\right) \left(m_B^2 \lambda N_2 + 24 \mkhs T_1
\right)
\right\} Im(C^*_T B)  \nonumber \\
&& + 4 m_B^3 \mlh
\left\{ 8 \lambda T_1 \mkhs + m_B^2 \lambda^2 N_2
+ 2 \lambda \left(\mkhs + \sh - 1\right) N_1
\right\} Im(C^*_T C) + {3 \over 2} m_B^2 \sh \lambda Im(K^* M)
\nonumber \\
&& + 3 m_B^3 \mlh  \lambda
\left\{ m_B^2 \left( \sh Im(K^* H) + \left(1 - \sh - \mkhs\right)
Im(K^* G) \right) + Im(K^* F)
\right\}  \nonumber \\
&& + 16 m_B \mlh \lambda
\left\{ m_B^2 \left( 8 \mkhs T_1 Im(C^*_{TE} G)
+ N_2 \left( m_B^2 \lambda Im(C^*_{TE} G) - \left(\mkhs + \sh -
1\right)\right) N_2 \right)  \right. \nonumber \\
&& \left. - 2 \left( Im(C^*_{TE} F) - m_B^2 \left(\mkhs + \sh - 1
\right) Im(C^*_{TE} G) \right)
\right\}
+ 16 m_B^2 \sh
\left\{ 4 \left(\lambda + 12 \sh \mkhs\right) N_1^2
+ m_B^4 \lambda^2 \right. \nonumber \\
&& \left. + 192 \mkhs T_1^2 + 4 m_B^2 \lambda \left(\mkhs +
\sh - 1\right) N_1 N_2 + 96 \left(\mkhs + \sh - 1\right) T_1 N_1
+ 16 \lambda m_B^2 T_1 N_2
\right\} \nonumber \\
&& \times Im(C^*_{TE} C_T) \Bigg] ,
\label{pol-double:nt} \\
%%%%%%%%%%%%%%%%%%%%%%%%%%%%%%%%%%
{\cal P}_{TL} &=& \frac{1}{\Delta}
\frac{m_B^2 \pi}{\sh \mkhs}  \sqrt{\frac{\lambda}{\sh}}\sqrt{1 -
\frac{4 \mlhs}{\sh}}
\Bigg[
 2 m_B^2 \mkhs \mlh \sh Re(A^* F + B^* E + 8 T_1 C_T^* F)
+ \frac{m_B \sh}{2} \left\{- m_B^2 \lambda
\left(Re(K^* C) \right. \right. \nonumber \\
&& \left. \left. - Re(M^* G) \right) - \left(1 - \sh - \mkhs \right)
\left(Re(B^* K) + Re(M^* F) \right) \right\}
- 8 m_B^3 \mkhs \sh \left\{ \sh N_1 + \left(\mkhs + \sh - 1\right) T_1
\right\}   \nonumber \\
&& \times \left( Re(A^* C_T) - 2 Re(C_T^* E) \right)
 + 32 m_B \sh T_1 Re(B^* C_{TE})
+ \mlhs \sh
\left\{ \left(\mkhs + \sh - 1\right) \left(|F|^2 + m_B^4  \lambda |G|^2
\right)
\right. \nonumber \\
&& \left. - 2 m_B^2 \lambda Re(F^* G)
+ m_B^2 \left(\mkhs + \sh - 1\right) \sh Re(F^* H)
- m_B^4 \sh \lambda Re(G^* H)
\right\}
+ 8 m_B^2 \mlh \sh \left\{ m_B^2 \lambda N_2 \right. \nonumber \\
&& \left. + 2 \left(\mkhs + \sh - 1
\right) N_1 + 8 \mkhs T_1 \right\} Re(K^* C_{TE})
- 256 m_B^2 \mlh \left\{ \sh N_1 + \left(\mkhs + \sh - 1\right) T_1
\right\}    \nonumber \\
&& \left( |C_T|^2 + 4 |C_{TE}|^2 \right) - 16 m_B^3 \sh \left\{\sh
N_1 + \left(\mkhs + \sh - 1\right) T_1 \right\} Re(C_{TE}^* E)
\Bigg] ,
\label{pol-double:tl}  \\
%%%%%%%%%%%%%%%%%%%%%%%%%%%%%%%%
{\cal P}_{TN} &=& \frac{4}{\Delta} \frac{m_B^2}{3 \mkhs} \sqrt{1 -
\frac{4 \mlhs}{\sh}}
\Bigg[ 3 m_B^4 \sh \lambda Im(E^* A)
+ 32 m_B^3 \mlh \lambda \mkhs T_1
\left(- 2 Im(A^* C_{TE}) + Im(E^* C_T) \right)  \nonumber \\
&& + \lambda
\left\{ \left(\mkhs + \sh - 1\right) \left( m_B^2 Im(G^* B) - Im(C^*
F) \right) - Im(F^* B) + m_B^2 \lambda Im(C^* G)
\right\} \nonumber \\
&& + 4 m_B \mlh
\left\{ 2 \left(\lambda + 12 \sh \mkhs\right) N_1
+ \left(\mkhs + \sh - 1\right) \left(m_B^2 \lambda N_2 + 24 \mkhs T_1
\right)
\right\} Im(C^*_T B)  \nonumber \\
&& + 4 m_B^3 \mlh
\left\{ 8 \lambda T_1 \mkhs + m_B^2 \lambda^2 N_2
+ 2 \lambda \left(\mkhs + \sh - 1\right) N_1
\right\} Im(C^*_T C) + {3 \over 2} m_B^2 \sh \lambda Im(K^* M)
\nonumber \\
&& + 3 m_B^3 \mlh  \lambda
\left\{ m_B^2 \left( \sh Im(K^* H) + \left(1 - \sh - \mkhs\right)
Im(K^* G) \right) + Im(K^* F)
\right\}  \nonumber \\
&& + 48 m_B \mlh \lambda
\left\{ m_B^2 \left( 8 \mkhs T_1 Im(C^*_{TE} G)
+ N_2 \left( m_B^2 \lambda Im(C^*_{TE} G) - \left(\mkhs + \sh -
1\right)\right) N_2 \right)  \right. \nonumber \\
&& \left. - 2 \left( Im(C^*_{TE} F) - m_B^2 \left(\mkhs + \sh - 1
\right) Im(C^*_{TE} G) \right)
\right\}
+ 16 m_B^2 \sh
\left\{ 4 \left(\lambda + 12 \sh \mkhs\right) N_1^2
+ m_B^4 \lambda^2 \right. \nonumber \\
&& \left. + 192 \mkhs T_1^2 + 4 m_B^2 \lambda \left(\mkhs +
\sh - 1\right) N_1 N_2 + 96 \left(\mkhs + \sh - 1\right) T_1 N_1
+ 16 \lambda m_B^2 T_1 N_2
\right\} \nonumber \\
&& \times Im(C^*_{TE} C_T) \Bigg] ,
\label{pol-double:tn}  \\
%%%%%%%%%%%%%%%%%%%%%%%%%%%%%%%%%
{\cal P}_{TT} &=& \frac{2}{3 \Delta} m_B^2
\Bigg[ \lambda \left(\sh + 4 \mlhs\right) m_B^4 |A|^2
+ \frac{1}{\sh} \left\{ \lambda \sh - 2 \left(\lambda + 12 \sh
\mkhs\right) \right\} |B|^2
+ \frac{\left(2 \mlhs - \sh\right)}{\sh \mkhs} \lambda m_B^2
\left\{ \lambda |C|^2 \right. \nonumber \\
&& \left. - 2 \left(\mkhs + \sh - 1\right) Re(B^* C) \right\}
+ 128 m_B^3 \mlh \lambda T_1 Re(A^* C_T)
+  m_B^2 \lambda \left(\sh - 4 \mlhs\right)
\left( - |E|^2  \right. \nonumber \\
&& \left. + \frac{3}{2 \mkhs} |K|^2 \right)
+ 16 m_B \frac{\mlh}{\mkhs}
\left\{ 2 \left(\lambda - 12 \sh \mkhs\right) N_1
 + \left(\mkhs + \sh - 1\right) \left( m_B^2 \lambda N_2
 - 24 \mkhs T_1\right) \right\} \nonumber \\
&& \times Re(B^* C_{TE})
- 32 m_B^3 \frac{\mlh \lambda }{\mkhs}
\left\{ 8 \mkhs T_1 + m_B^2 \lambda N_2 - 2 \lambda \left(\mkhs + \sh
- 1 \right) N_1 \right\} Re(C^* C_{TE})  \nonumber \\
&& + m_B^4 \frac{\lambda}{\mkhs \sh} \left\{ \lambda \sh - 2 \mlhs
\left(5 \lambda + 12 \sh \mkhs\right) \right\} |G|^2
+ \frac{\left(\sh - 10 \mlhs\right)\lambda}{\sh \mkhs}
\left\{ |F|^2 - 2 m_B^4\left(\mkhs + \sh - 1\right) Re(F^* G) \right\}
\nonumber \\
&& - 12 m_B^2 \frac{\mlhs \lambda}{\mkhs} Re(F^* H)
+ 6 m_B \frac{\mlh \lambda}{\mkhs}
\left\{ m_B^2 \left( \left(\mkhs + \sh - 1\right) Re(M^* G) - \sh
Re(H^* M) \right) - Re(M^* F)
\right\}   \nonumber \\
&& - {3 \over 2} m_B^2 \frac{\lambda \sh}{\mkhs} |M|^2
+ 8 \frac{m_B^2}{\sh \mkhs}
\left\{ \left(4 \mlhs - \sh\right)
\left( \left(\lambda + 12 \sh \mkhs\right) \sh N_1^2
+ \lambda^2 m_B^4 N_2^2 + 16 \sh \mkhs m_B^2 \lambda \right)
\right. \nonumber \\
&& \left. + 4 \left(4 \mlhs - \sh\right) \left(\mkhs + \sh - 1\right)
\sh  \left( \lambda m_B^2 N_1 N_2 + 24 \mkhs N_1 T_1 \right)
+ 64 \mkhs \left( 4 \mlhs \lambda - 3 \mkhs \sh^2 \right) T_1^2
\right\} \nonumber \\
&& \times |C_T|^2
+ 32 \frac{m_B^2}{\sh \mkhs}
\left\{ 4 \left( \left(\lambda + 12 \sh \mkhs\right) \sh - 8 \mlhs
\lambda \right) \sh N_1^2
-  \sh \lambda m_B^2 N_2 \left(8 \mlhs - \sh\right)
\left(\lambda N_2 + 2 \mkhs T_1 \right) \right.  \nonumber \\
&& \left. - 4 m_B^2 \left(8 \mlhs - \sh\right) \left(\mkhs + \sh -
1\right) \sh \lambda N_1 N_2
+ 64 \mkhs \left( 4 \lambda \mlhs + 3 \mkhs \sh^2\right) T_1^2
+ 96 \left(\mkhs + \sh - 1 \right) \mkhs \right. \nonumber \\
&& \left. + \sh^2 N_1 T_1 \right\} |C_{TE}|^2 \Bigg] .
\label{pol-double:tt}
\end{eqnarray}

%%%%%%%%%%%%%%%%%%%%%%%%%%%%%%%%%%
%  Section : 4

\section{Numerical analysis, Results and Discussion \label{section:4}}

In this final section we shall present the results of our
numerical analysis. As such, the input parameters which we have
used, in order to calculate the various Wilson coefficients
defined in Eqn.(\ref{effH}), are listed in Appendix
\ref{inputpara}. The value of $C_7$ is fixed by the observation of $b
\to s \gamma$. Note that this observation fixes the magnitude and not
the sign of $C_7$, we have therefore choosen the SM predicted value
$C_7 = - 0.313$. For $C_{10}$ we have used the SM value $C_{10} = -
4.997$. Regarding the value of $C_9^{eff}$ this receives both short
and long-distance contributions. The long-distance contributions are
the result of $c \bar{c}$ intermediate states, such as resonances of
$J/\Psi$. For our analysis we have used $C_9^{eff} = C_9 + Y(\hat{s})$
where $C_9$ corresponds to the short distance contribution, which we
have taken to have SM value $C_9 = 4.334$. $Y(\hat{s})$ represents the
${\cal O}(\alpha_s)$ corrections coming from the operators $O_1 - O_6$
as given in Kruger \& Sehgal \cite{Kruger:1996cv}. In our present
analysis we have confined ourselves only to short distance
contributions. The form factor definitions which we have used in
describing the hadronic transitions are listed in Appendix
\ref{formfactor}. In our effective Hamiltonian there are in all 12
Wilson coefficients. Of these $C_{SL}$ and $C_{BR}$ can be related to
the SM Wilson $C_7^{eff}$ by; 
\begin{equation} 
 C_{SL} = - 2 m_s C_7^{eff} ~~,~~ C_{BR} = - 2 m_b C_7^{eff} .
\label{sec:4:eq:1}
\end{equation}
Among the vector type coefficients $C_{LL}$, $C_{LR}^{tot}$, $C_{RL}$
and $C_{RR}$ two  
of them, namely $C_{LL}$ and $C_{LR}^{tot}$, are already defined in
terms of the SM Wilson coefficients $(C_9^{eff} - C_{10})$ and
$(C_9^{eff} + C_{10})$  respectively. The remaining vector type
interaction coefficients $C_{RL}$ and $C_{RR}$ are taken to be free
parameters. The coefficients of the scalar type interactions, namely
$C_{LRLR}$, $C_{RLLR}$, $C_{LRRL}$ and $C_{RLRL}$, and the tensor type
interactions, $C_T$ and $C_{TE}$, are also taken to be input
parameters. 

%%%%%%%%%%%%%%%%%%%%%%%%%%%%%%%%%%%%%%%%
\FIGURE[ht]{
\epsfig{file=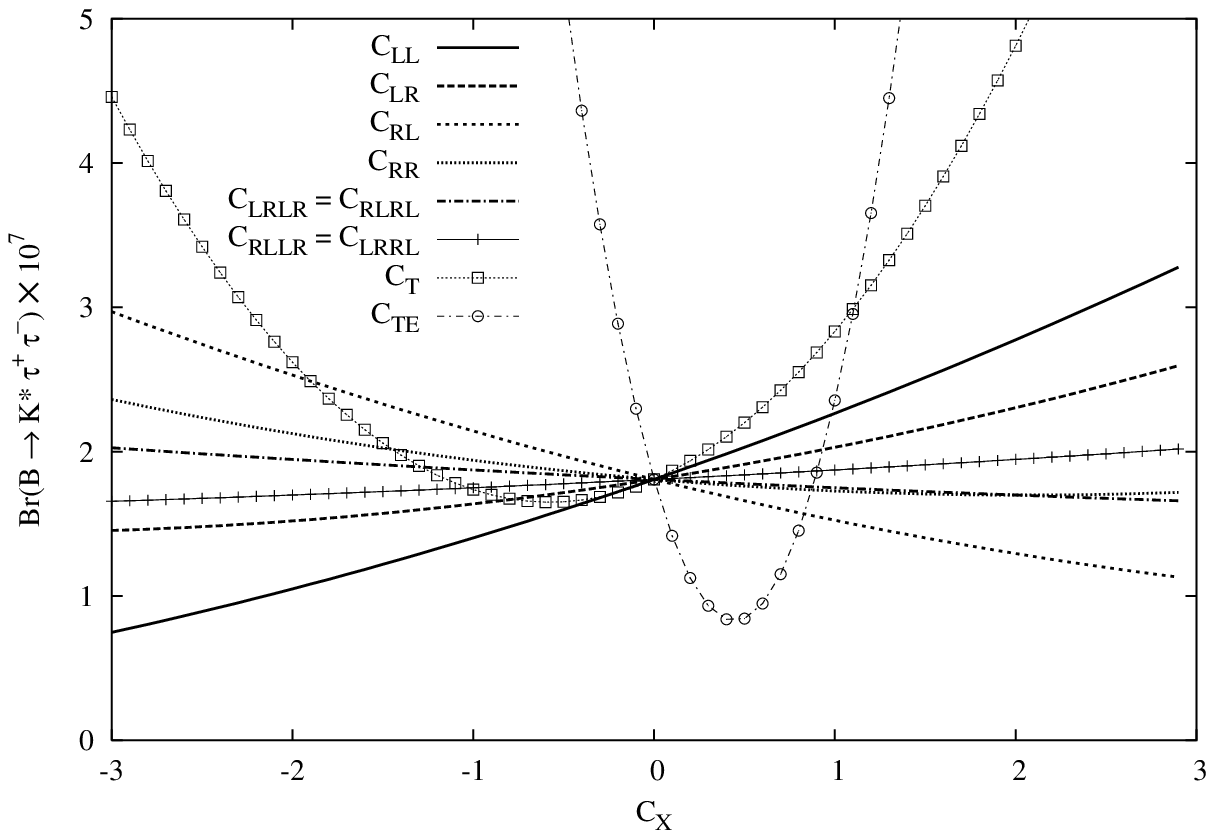,width=.8\textwidth,height=2.6in} \caption{The
branching ratio, $Br \left( B \to K^* \tau^+ \tau^- \right)$, as a
function of the various Wilson coefficients.} \label{fig:1}
}
%%%%%%%%%%%%%%%%%%%%%%%%%%%%%%%%%%%%%%%%%
%%%%%%%%%%%%%%%%%%%%%%%%%%%%%%%%%%%%%%%%
\FIGURE[ht]{
\epsfig{file=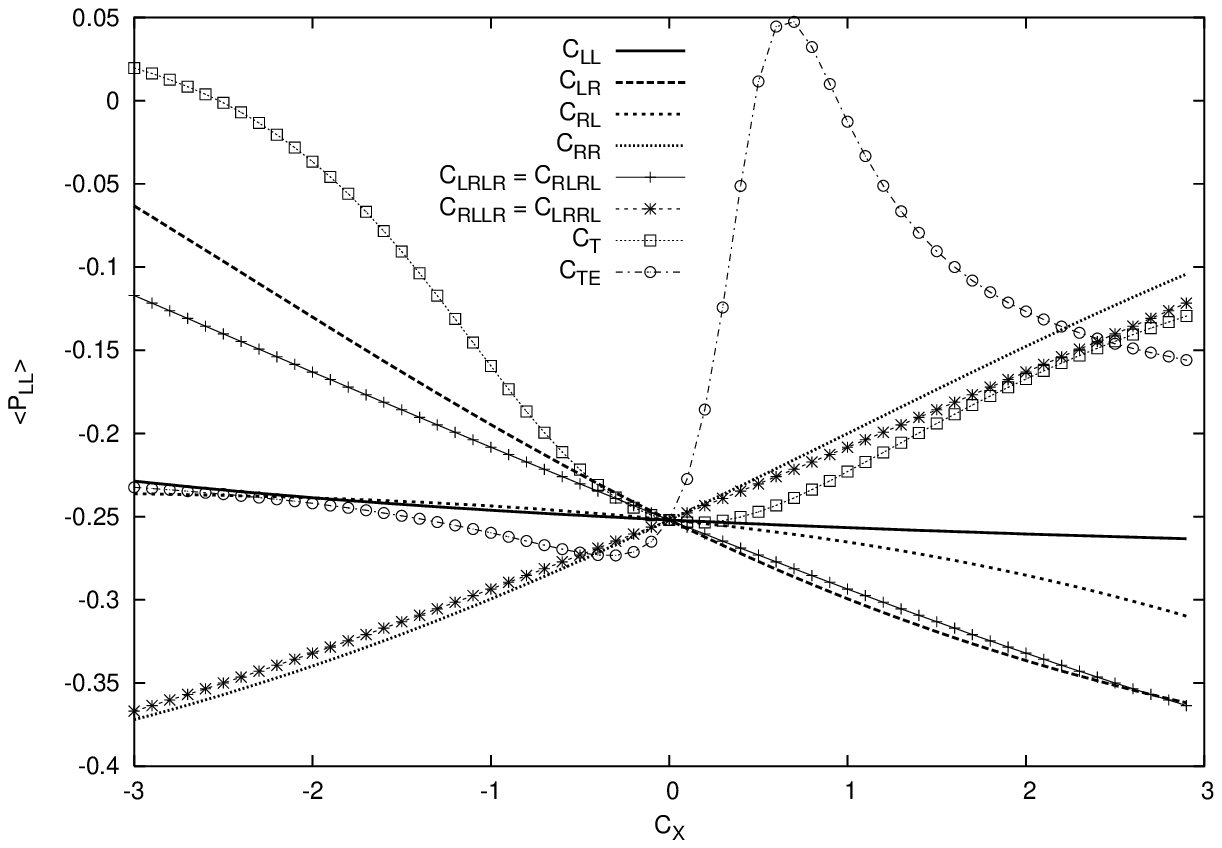,width=.8\textwidth,height=2.6in}
\caption{The double polarization asymmetry, ${\cal P}_{LL}$, as a
function of the various Wilson coefficients, where both $\tau$
leptons are longitudinally polarized.} \label{fig:2}
}
%%%%%%%%%%%%%%%%%%%%%%%%%%%%%%%%%%%%%%%%%

\par As already discussed in the introduction we shall explain the
$B \to \pi \pi$ and $B \to K \pi$ puzzle as resulting from a large
phase in the electroweak penguin diagrams as has been proposed in
\cite{Buras:2000gc}. This suggestion was initially made by Buras {\em
et al.} some time back \cite{Buras:2000gc} and has lately been revived
by many other groups \cite{Yoshikawa:2003hb}. Recently the
implications of this suggestion on a variety of hadronic, leptonic and
semi-leptonic processes has been studied. Note that, as emphasized in
our earlier work on the inclusive decay mode $B \to X_s \ell^+ \ell^-$
\cite{RaiChoudhury:2004pw}, the polarization asymmetries could also
significantly deviate from their SM values if there was a large phase
in the electroweak penguins. This type of study has also been carried
out by Aliev {\em et al.} \cite{Kruger:2000zg} where they attempted to
estimate the variation in the single polarization asymmetries for the
exclusive process $B \to K^* \ell^+ \ell^-$, where the Wilsons had
some extra phase. Aliev {\em et al.} in their study of the single
lepton polarization asymmetries in the exclusive process $B \to K^*
\ell^+ \ell^-$ also emphasized the importance of the tensorial
interactions on various asymmetries \cite{Aliev:2000jx}. They
concluded that single polarization asymmetries are very sensitive to
scalar and tensor type interactions. In our earlier work
\cite{Choudhury:2003mi} we demonstrated the supersymmetric effects on
various double polarization asymmetries in $B \to K^* \ell^+ \ell^-$,
where supersymmetry predicts the existence of scalar and pseudo-scalar
operators in the large $tan\beta$ region\footnote{$tan\beta$ is the
ratio of the vev's of two Higgs bosons} \cite{Choudhury:1998ze}. 
However, in this previous study we did not include the tensorial
structures. In this current work we will use the most general form of
the effective Hamiltonian to study the effects on various polarization
asymmetries. We shall also include the extra possible phase from the
electroweak penguin sector. 

%%%%%%%%%%%%%%%%%%%%%%%%%%%%%%%%%%%%%%%%
\FIGURE[ht]{
\epsfig{file=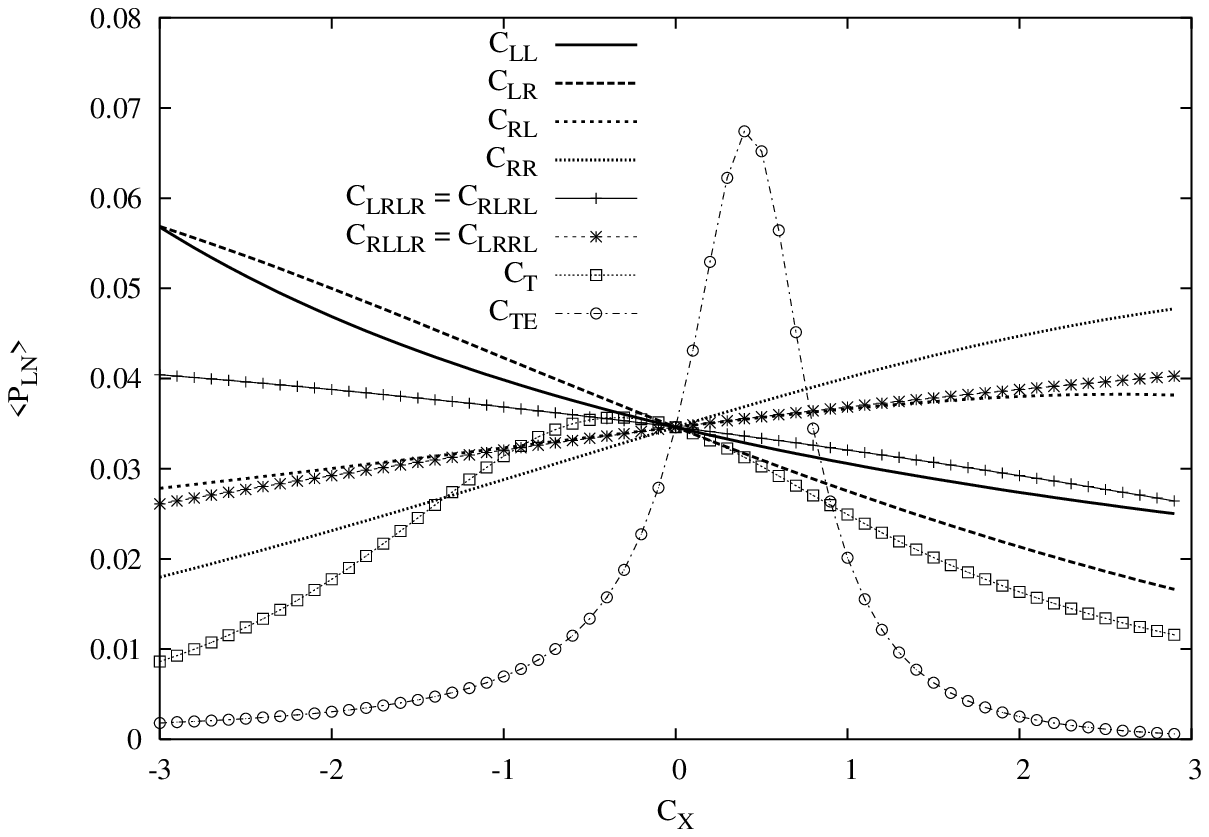,width=.8\textwidth,height=2.6in}
\caption{The same as Figure (\ref{fig:2}), but for $\ell^-$
longitudinal and $\ell^+$ normal.} \label{fig:3}
}
%%%%%%%%%%%%%%%%%%%%%%%%%%%%%%%%%%%%%%%%%
%%%%%%%%%%%%%%%%%%%%%%%%%%%%%%%%%%%%%%%%
\FIGURE[ht]{
\epsfig{file=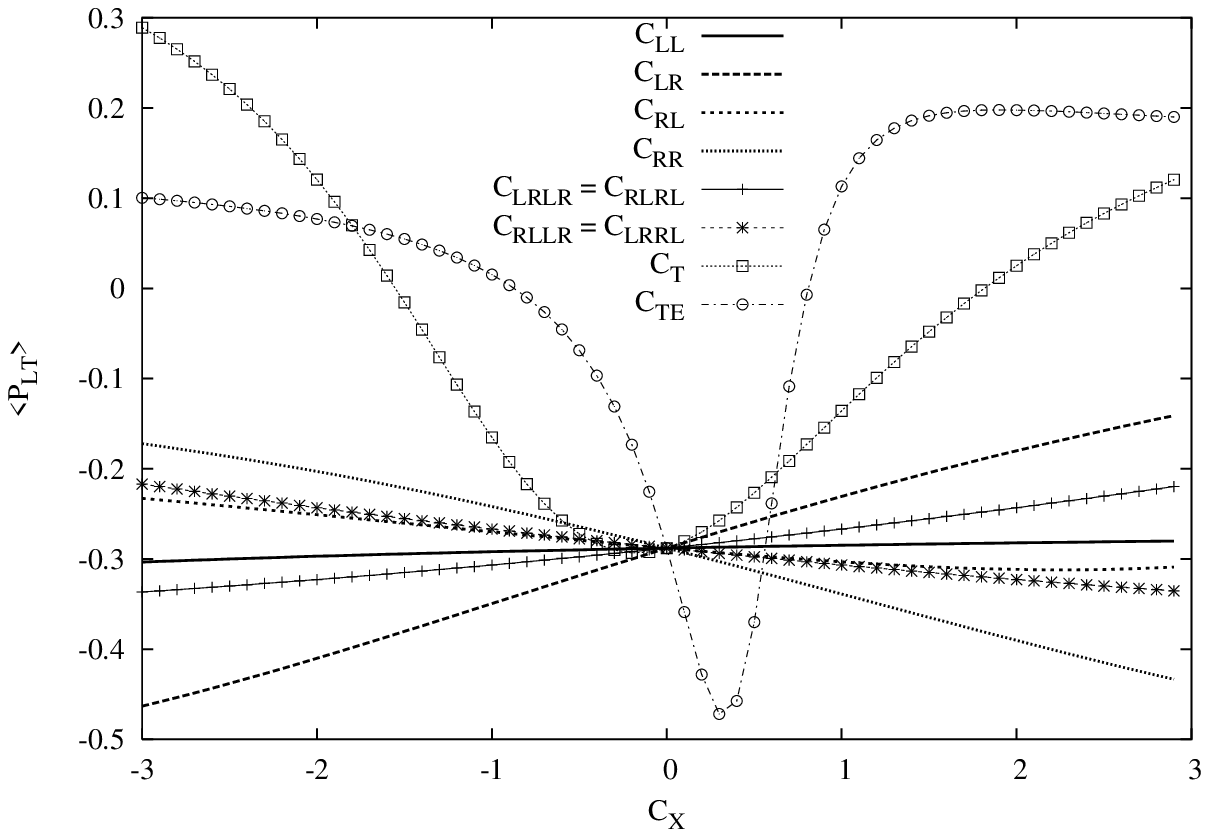,width=.8\textwidth,height=2.6in}
\caption{The same as Figure (\ref{fig:2}), but for $\ell^-$
longitudinal and $\ell^+$ transverse.} \label{fig:4}
}
%%%%%%%%%%%%%%%%%%%%%%%%%%%%%%%%%%%%%%%%%
\par Note that in this paper these additional effects from the
electroweak penguin sector, which can give effective structures
similar to those given in Eqn.(\ref{effH}) with coefficients
$C_{LL}$, $C_{LR}$, $C_{RL}$ and $C_{RR}$, will all be given an
additional phase. As already stated in section \ref{section:2} two
of these coefficients, namely $C_{LL}$ and $C_{LR}$, can be
parameterized in terms of $C_9$ and $C_{10}$. Therefore we shall
only consider effects of a new phase in the $C_{10}$, $C_{RL}$ and
$C_{RR}$ coefficients. For this purpose we will parameterize these
coefficients as;
\begin{eqnarray}
C_{10} &=& |C_{10}| e^{i \phi_{10}} ,
\label{sec4:eq:1}  \\
C_{RL} &=& |C_{RL}| e^{i \phi_{RL}} ,
\label{sec4:eq:2}  \\
C_{RR} &=& |C_{RR}| e^{i \phi_{RR}} .
\label{sec4:eq:3}
\end{eqnarray}
%%%%%%%%%%%%%%%%%%%%%%%%%%%%%%%%%%%%%%%%
\FIGURE[ht]{
\epsfig{file=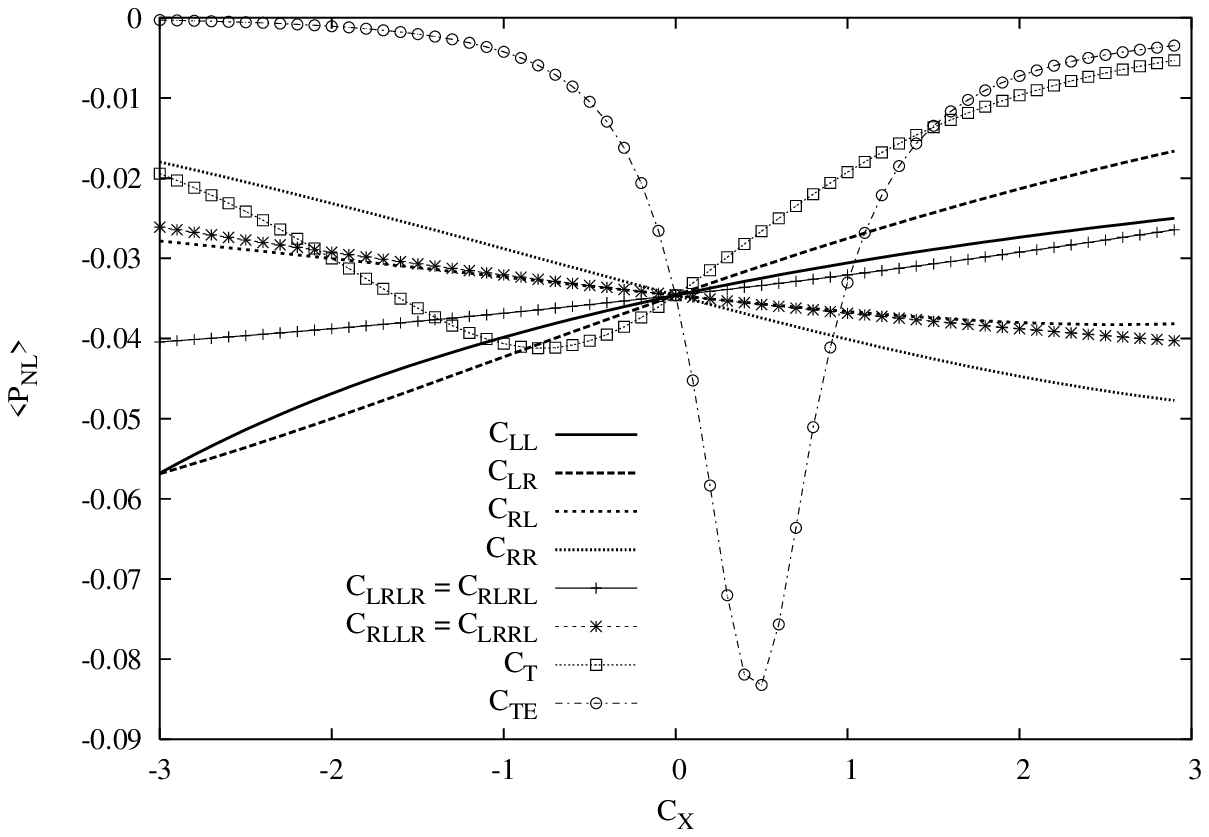,width=.8\textwidth,height=2.6in}
\caption{The same as Figure (\ref{fig:2}), but for $\ell^-$ normal
and $\ell^+$ longitudinal.} \label{fig:5}
}
%%%%%%%%%%%%%%%%%%%%%%%%%%%%%%%%%%%%%%%%%
%%%%%%%%%%%%%%%%%%%%%%%%%%%%%%%%%%%%%%%%
\FIGURE[ht]{
\epsfig{file=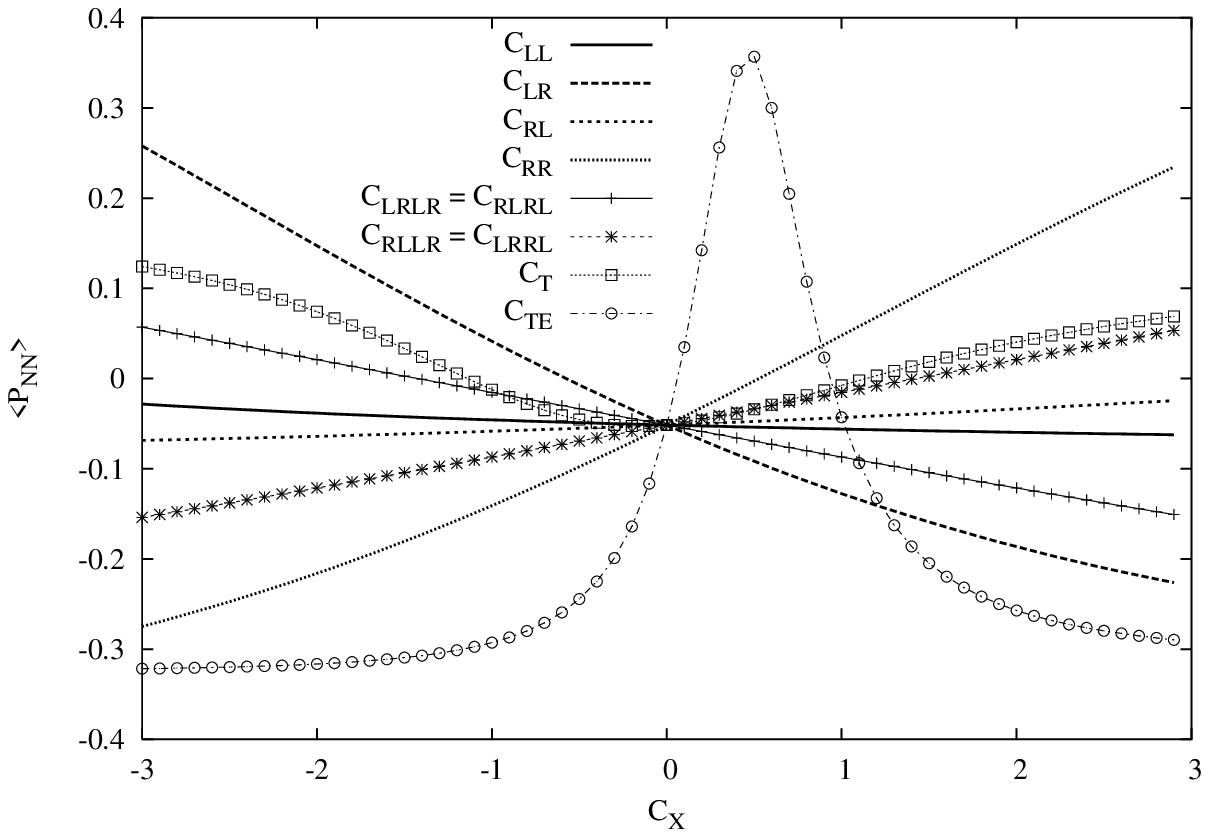,width=.8\textwidth,height=2.6in}
\caption{The same as Figure (\ref{fig:2}), but for both leptons
polarized in the normal direction.} \label{fig:6}
}
%%%%%%%%%%%%%%%%%%%%%%%%%%%%%%%%%%%%%%%%%

\par As the majority of our results involve the polarization
asymmetries, listed in the previous section, which are dependent
on the scaled invariant mass ($\hat{s}$), it is experimentally
useful to consider the averaged values of these asymmetries.
Therefore we shall present only the averaged values of the
polarization asymmetries in our results using the averaging
procedure defined as;
\begin{equation}
\langle P_i \rangle = \frac{\displaystyle  \int^{(m_B -
m_{K^*})^2}_{4 m_\ell^2} P_i \frac{d \Gamma}{d s} ds}
{\displaystyle \int^{(m_B - m_{K^*})^2}_{4 m_\ell^2} \frac{d
\Gamma}{d s} ds} . \label{sec:4:eq:3}
\end{equation}

\par Our results are presented in a series of figures commencing with
Figure (\ref{fig:1}) where we have plotted the branching ratio of $B
\to K^* \tau^+ \tau^-$ as a function of the various Wilson
coefficients. In this plot we have constrained the value of the
branching ratio to have an upper bound $Br(B \to K^* \tau^+ \tau^-)
\le 5 \times 10^{-7}$. As can be seen from this graph the largest
variation in the branching ratio corresponds to tensorial type
interactions. 

\par Figures (\ref{fig:2})-(\ref{fig:10}) represent the various double
polarization asymmetries plotted as functions of the various
Wilson coefficients. In these plots we have assumed that all the
Wilson coefficients are real. In all cases we have varied the
Wilson coefficients over a range of -3 to 3. It is also apparent
that, as in the case of the branching ratio, the greatest variation
of the various double polarization asymmetries corresponds to the
tensorial type interactions. From the graphs of the polarization
asymmetries we can also see substantial variations for various
other values of the Wilson coefficients. The major change is that
produced in the plots of $<{\cal P}_{LL}>$, $<{\cal P}_{LT}>$, $<{\cal
P}_{NN}>$, $<{\cal P}_{NT}>$, $<{\cal P}_{TL}>$, $<{\cal P}_{TN}>$ and
$<{\cal P}_{TT}>$ where the respective asymmetry can even change
sign for certain values of the Wilson coefficients! Note that in
these plots we have only shown those asymmetries which are larger
than $10^{-3}$. As such the variations of $C_{LL}$, $C_{RL}$ and
$C_{LR}$, in the respective asymmetries for Figures (\ref{fig:7})
and (\ref{fig:9}), are not shown.
%%%%%%%%%%%%%%%%%%%%%%%%%%%%%%%%%%%%%%%%
\FIGURE[ht]{
\epsfig{file=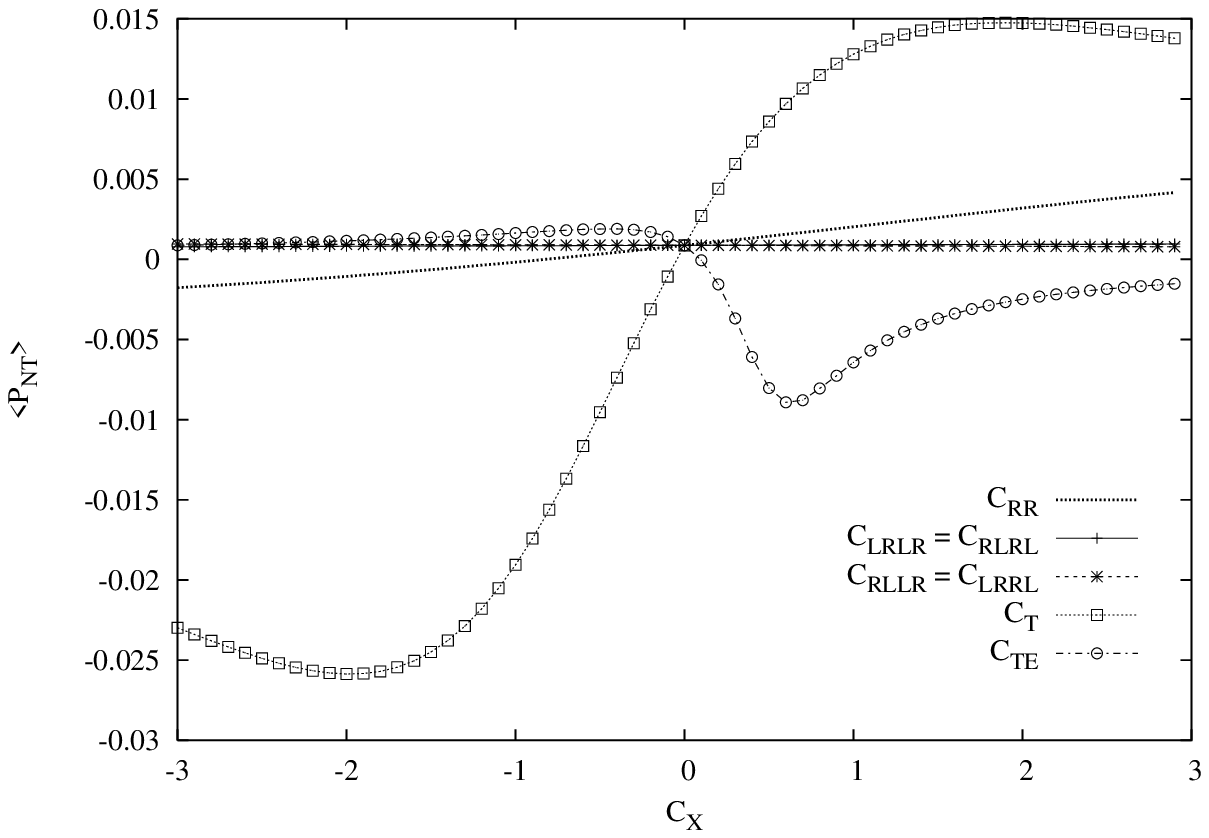,width=.8\textwidth,height=2.6in}
\caption{The same as Figure (\ref{fig:2}), but for $\ell^-$ normal
and $\ell^+$ transverse.} \label{fig:7}
}
\FIGURE[ht]{
\epsfig{file=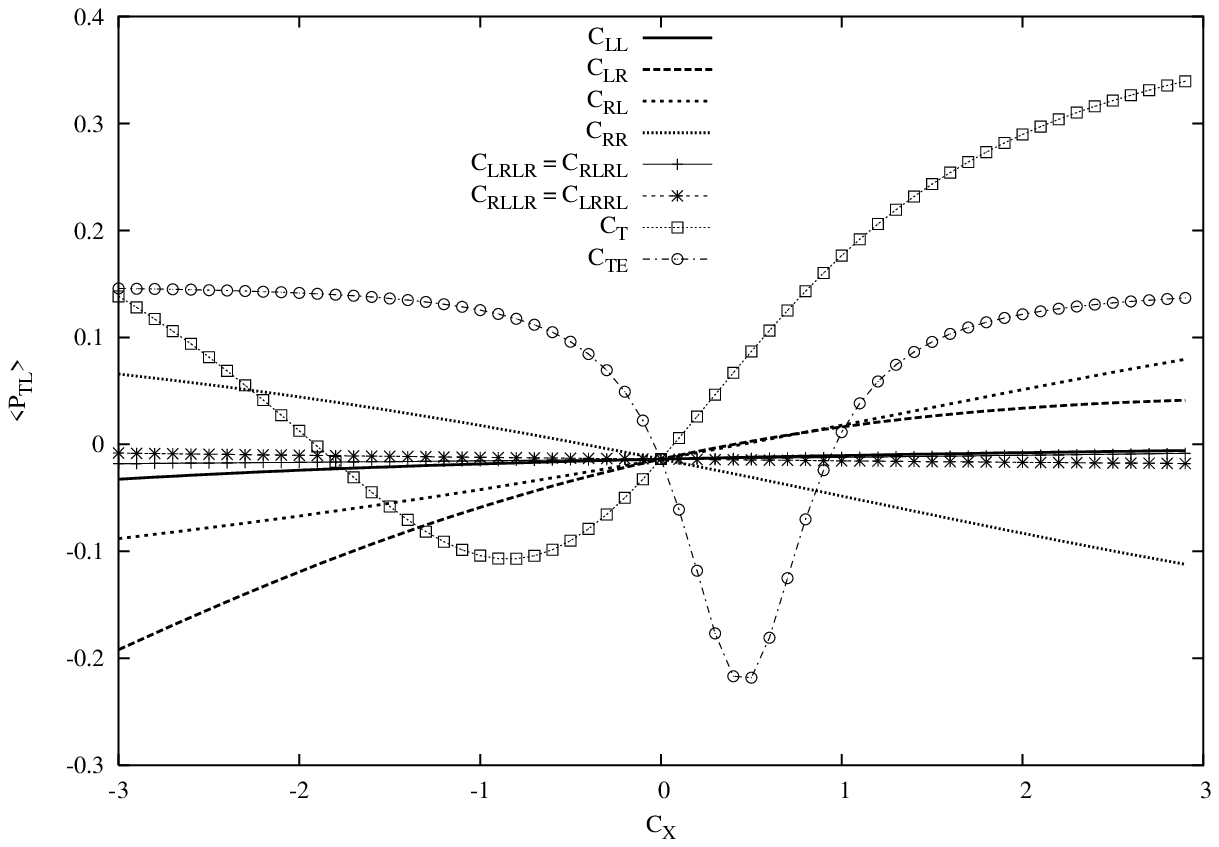,width=.8\textwidth,height=2.6in}
\caption{The same as Figure (\ref{fig:2}), but for $\ell^-$
transverse and $\ell^+$ longitudinal.} \label{fig:8}
}
%%%%%%%%%%%%%%%%%%%%%%%%%%%%%%%%%%%%%%%%%
\par Note in particular that the $<{\cal P}_{LL}>$ in Figure
(\ref{fig:2}) shows substantial dependence on the $C_{LR}$,
$C_{LRLR}$, $C_{RLLR}$, $C_{RR}$, $C_T$ and $C_{TE}$ coefficients in
which the magnitude of the asymmetry can change by more than 100\%. Of
major significance is that the tensorial operators can even change the
sign of this asymmetry. A very similar sort of behaviour is
exhibited by these Wilson coefficients for $<{\cal P}_{LN}>$, $<{\cal
P}_{NL}>$ and $<{\cal P}_{NT}>$, except here in the case of $<{\cal
P}_{LN}>$ and $<{\cal P}_{NL}>$ the sign of the asymmetry does not
change. We can also see in Figure (\ref{fig:6}) that for the case
of $<{\cal P}_{NN}>$ all the Wilson coefficients, with the exception
of $C_{LL}$ and $C_{RL}$, predict a sign change.
%%%%%%%%%%%%
% addition
\par These results prompt us to analyze the polarization
asymmetries for the case where the branching ratio of $B \to K^*
\tau^+ \tau^-$ remains close to the SM value. This sort of scenario
could tell us more about how the various Wilson coefficients affect 
various asymmetries (as they are different quadratic functions of the
Wilsons and hence carry independent sets of information). This
possibility has been presented in Figure (\ref{fig:10(a)}), where we
have also restricted the branching ratio to the range $1 \times
10^{-7} < Br(B \to K^* \tau^+ \tau^-) < 4 \times 10^{-7}$. From these
graphs we observe that there can be substantial variation in the
polarization asymmetries even if the branching ratio is not
substantially different from its SM value. As can be seen from Figure
(\ref{fig:10(a)}) all the polarization asymmetries shows substantial
variations. Some of the asymmetries, in particular $<P_{LL}>$,
$<P_{LT}>$, $<P_{NN}>$, $<P_{NT}>$, $<P_{TL}>$,  $<P_{TN}>$ and
$<P_{TT}>$ not only show variation in magnitude but even their sign
changes. Some asymmetries like $<P_{NN}>$, $<P_{TL}>$ and $<P_{TT}>$
can change by more than an order of magnitude, even if there is no
major change in the branching ratio. Note that all these asymmetries
are most sensitive to tensorial structures in the effective
Hamiltonian.  
%%%%%%%%%%%%

\par The next set of results we present includes the extra
phase in the Wilson coefficients, as stated in
Eqns.(\ref{sec4:eq:1}), (\ref{sec4:eq:2}) and (\ref{sec4:eq:3}).
In Figures (\ref{fig:11}) and (\ref{fig:12}) we have plotted
integrated polarization asymmetries as a function of the phase
$\phi_{10}$. In Figure (\ref{fig:11}) we have used the SM value
$C_{10} = 4.669$. Note that we have only shown those asymmetries which
vary with the inclusion of $\phi_{10}$. In Figure (\ref{fig:12}) we
have plotted the same variables but with an increase in the magnitude
of $C_{10}$, namely we have chosen $|C_{10}| = 9$. This value has been
chosen to correspond with the value calculated by Buras {\sl et al.}
\cite{Buras:2000gc} which they predict in order to solve the $B \to
\pi \pi$ and $B \to K \pi$ puzzle. They say that $C_{10}$ should be 
complex with a magnitude almost twice that of its SM value with a
phase which should make it almost imaginary. As can be seen in both
the Figures there is a substantial deviation as the phase,
$\phi_{10}$, is changed. 

%%%%%%%%%%%%%%%%%%%%%%%%%%%%%%%%%%%%%%%%
\FIGURE[ht]{
\epsfig{file=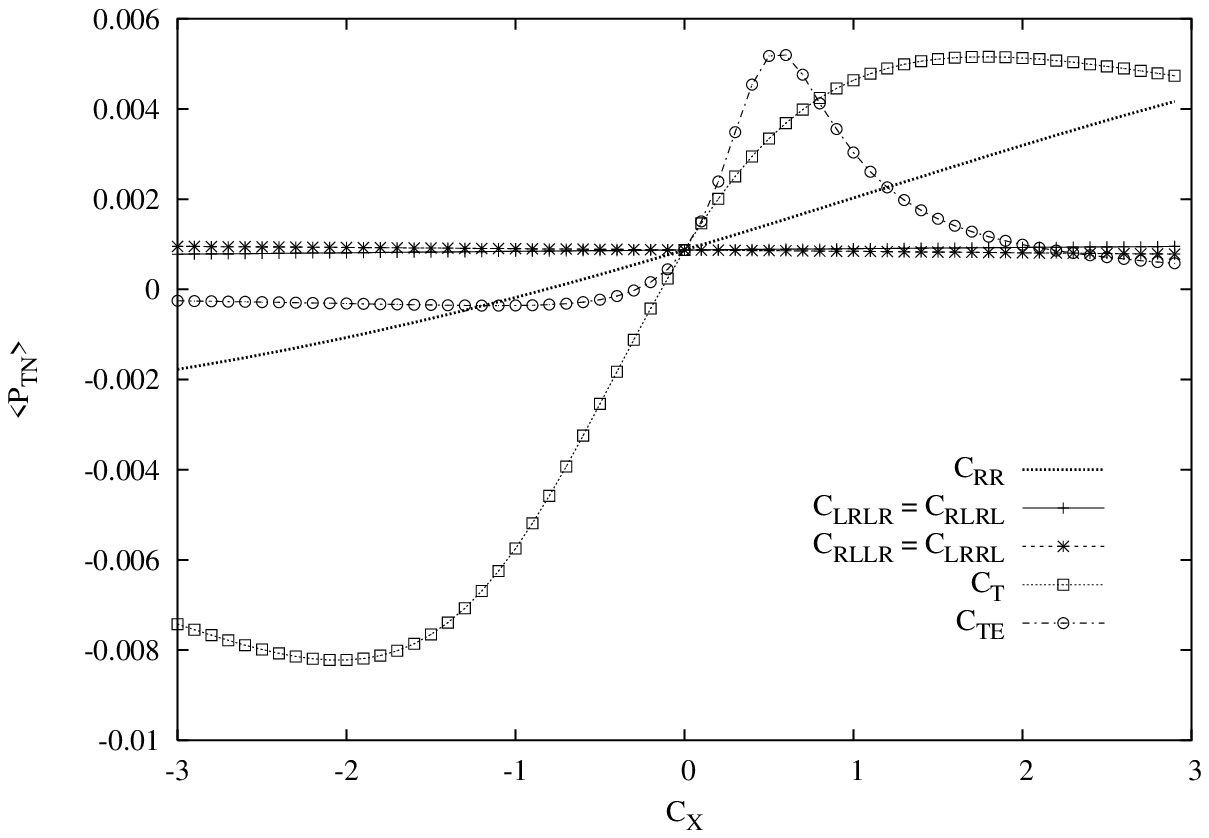,width=.8\textwidth,height=2.6in}
\caption{The same as Figure (\ref{fig:2}), but for $\ell^-$
transverse and $\ell^+$ normal.} \label{fig:9}
}
%%%%%%%%%%%%%%%%%%%%%%%%%%%%%%%%%%%%%%%%%
%%%%%%%%%%%%%%%%%%%%%%%%%%%%%%%%%%%%%%%%
\FIGURE[ht]{
\epsfig{file=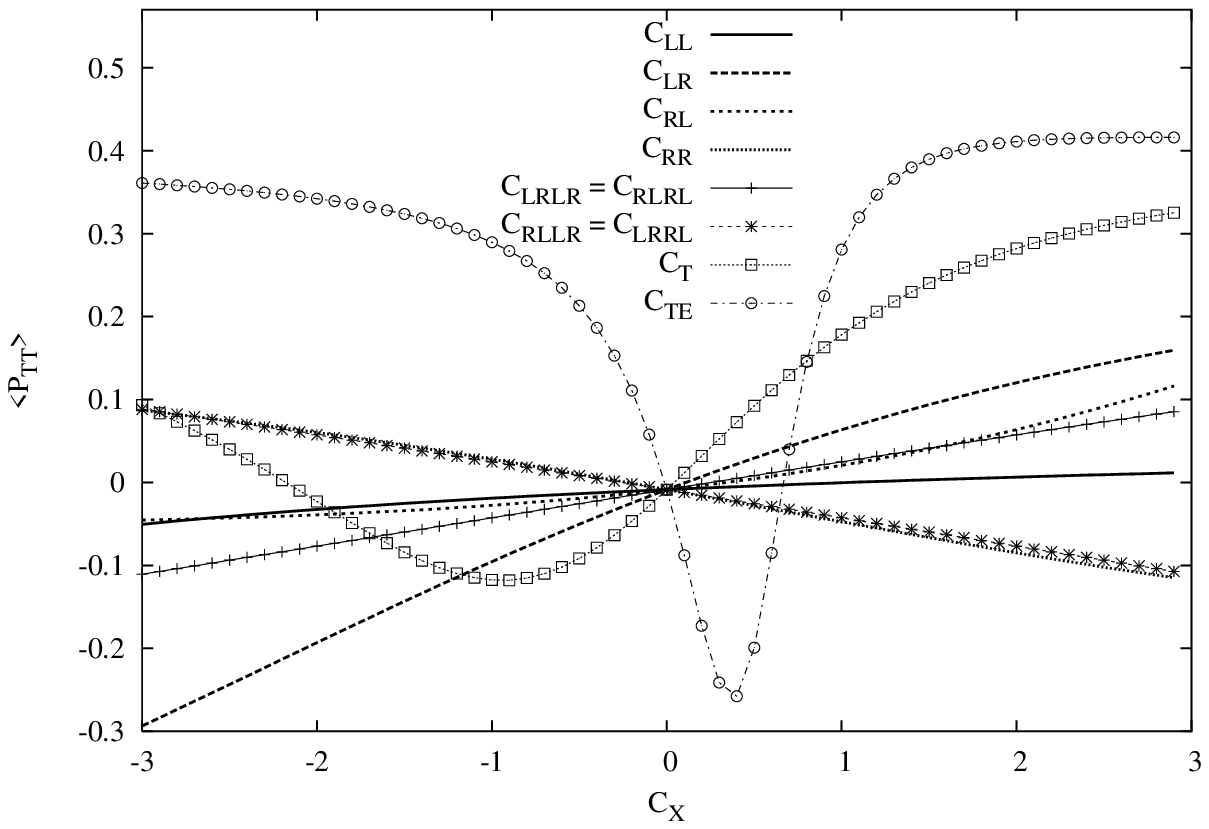,width=.8\textwidth,height=2.6in}
\caption{The same as Figure (\ref{fig:2}), but for both leptons
polarized in the transverse direction.} \label{fig:10}
}
%%%%%%%%%%%%%%%%%%%%%%%%%%%%%%%%%%%%%%%%%

\par In Figure (\ref{fig:13}) we have plotted the correlations of
various polarization asymmetries and the branching ratio of $B \to K^*
\tau^+ \tau^-$. In this plot we have varied the phase $\phi_{RL}$ in a
range of $0 \le \phi_{RL} \le \pi$. Finally in Figure (\ref{fig:14})
we have drawn the same sort of plot but for $C_{RR}$. As we can see
from these graphs some of the polarization asymmetries can change sign
as we vary the phase of the $C_{RL}$ and $C_{RR}$ Wilsons.
%%%%%%%%%
\par For the numerical analysis the central values of the form factors
given in Appendix \ref{formfactor} and \cite{Ali:1999mm} have been
used. These form factors have substantial uncertainties associated
with them but as can be seen from the various plots we have shown that
these polarization asymmetries have substantial variations even if we
restrict ourselves to a branching ratio near the SM value. In fact
some of the asymmetries can even change sign, as shown in Figure
(\ref{fig:10(a)}). Although changing the form factor definition will
change the quantitative nature of these asymmetries (and the branching
ratio) the qualitative nature of these observables would remain the
same (as the variations in the asymmetries is substantial and hence
even with uncertainties present in form factor definitions one can
still draw definite conclusions regarding the asymmetries). 
%%%%%%%%%%%%%
\par Finally, in order to measure these various polarization
asymmetries we must determine how many $B \bar{B}$ pairs are
required. Using the arguments given in
\cite{Choudhury:2003xg,Aliev:2004yf}, experimentally an observation of
the polarization asymmetry $<P>$ of a decay with branching ratio
${\cal B}$ at a $n\sigma$ level to the required number of events is
given by ({\sl i.e.} the number of $B \bar{B}$ pairs); 
\begin{eqnarray}
 N &=& \frac{n^2}{s_1 s_2 ~ {\cal B} <P_{ij}>} ~~~~~~ {\rm (for ~ double
~ polarization ~ asymmetries)} \nonumber \\
 N &=& \frac{n^2}{s ~ {\cal B} <P_i>} ~~~~~~ {\rm (for ~ single 
~ polarization ~ asymmetries)} 
\end{eqnarray}
In the above equation $s_1, s_2$ and $s$ are the efficiencies of
detecting the state of polarization of $\tau$ leptons. If we take
$s_1( = s_2 = s)$ to be 0.5 then the number of events required to
observe various asymmetries at the $3 \sigma$ level would
be\footnote{here we are taking the efficiency of detecting the
polarization state, including the efficiency of $\tau$ measurement and
detection of the polarization of $\tau$. If $\tau$ detection
efficiency is 80\% and efficiency of detection of its polarization
state is 60\% then the total efficiency of detecting $\tau$ in a
particular polarization state is $0.8 \times 0.6 \approx 0.5$.};
\begin{equation}
N ~=~ 
\cases{~ (2 \pm 1) \times 10^8 ~~~ 
               {\rm for}~<{\cal P}_L>,<{\cal P}_N>,<{\cal P}_T>,\cr
       ~ (8 \pm 3) \times 10^8 
    ~~~ {\rm for}~<{\cal P}_{LL}>, <{\cal P}_{LT}>, \cr
       ~ (5 \pm 3) \times 10^{9}
    ~~~ {\rm for}~<{\cal P}_{LN}>,<{\cal P}_{NL}>, \cr
       ~ (8 \pm 7.5) \times 10^9
    ~~~ {\rm for}~<{\cal P}_{NN}>,<{\cal P}_{TL}>,<{\cal P}_{TT}>, \cr
       ~ (2 \pm 1.5) \times 10^{10} 
    ~~~ {\rm for}~<{\cal P}_{NT}>,<{\cal P}_{TN}>. 
              \cr}
\end{equation}
The number of $B \bar{B}$ pairs required to observe these asymmetries
might not be produced at the present generation B-factories but future
factories like LHCb and Super-B would be able to measure these
asymmetries. 
%%%%%%%%%%%%%%%%%%%%%%%%%%%%%%%%%%%%%%%%%
% Figure
\FIGURE[ht]{
\epsfig{file=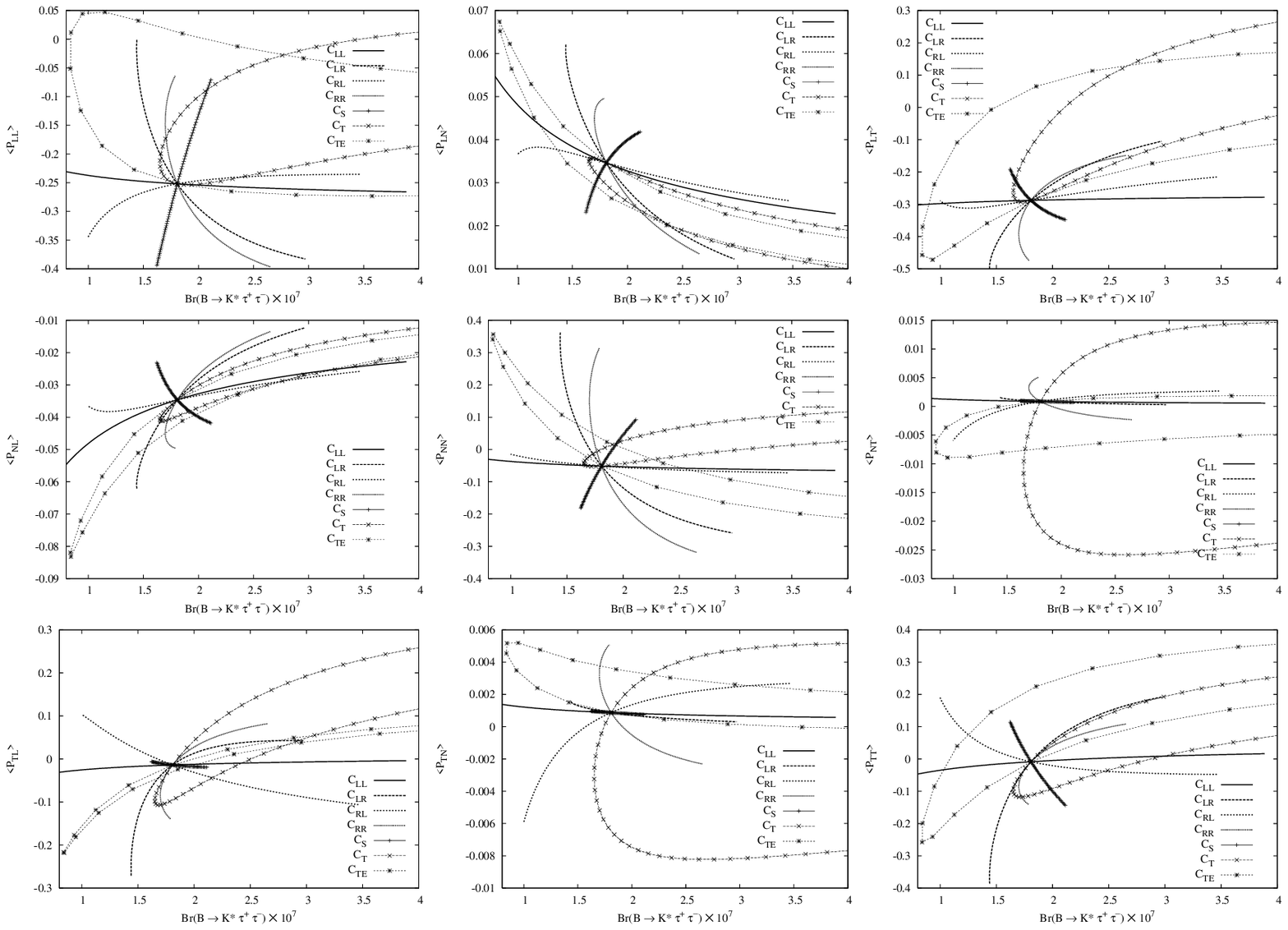,width=\textwidth} 
\caption{Plots of various integrated polarization asymmetries with
Branching ratio of $B \to K^* \tau^+ \tau^-$. In above figures 
$C_S = C_{LRLR} = C_{RLRL} = C_{RLLR} = C_{LRRL}$.} 
\label{fig:10(a)}
}
%%%%%%%%%%%%%%%%%%%%%%%%%%%%%%%%%%%%%%%%%

\par The study of polarization asymmetries in $B \to K^* \ell^+
\ell^-$ has also been done within the model independent framework by Aliev
{\sl et al.} \cite{Aliev:2000jx}. Their study of the single lepton
polarization asymmetries was done with real valued Wilsons. Our
results agree with those they obtained with the exception of
typographical errors in their expression of the normal polarization
asymmetries ${\cal P}_N^\pm$.  

\par The polarization asymmetries provide us a large number of
observables which would be very useful in determining the structure of
the effective Hamiltonian; which in turn could help us in discovering
the structure of the underlying physics. The most general effective
Hamiltonain for transitions based on $b \to s(d) \ell^+ \ell^-$ quark
level transition has twelve Wilsons. If we consider all these Wilsons
to be complex valued, this would result in 24 parameters and we would
need at least 24 observables in order to fix all these parameters. Of
these the measurement of $b \to s \gamma$ can fix one of them, namely
the magnitude of $C_7$. The observables which are presently at our
disposal are the branching ratio and the FB asymmetry. Along with
these two one can construct six single lepton polarization asymmetries
(three for each of the leptons) in addition one can have nine more
double polarization asymmetries. This would still leaves us with six
more unconstrained parameters. This number can be further restricted
if we also construct the polarized FB asymmetries (which would gives
us 15 more observables, namely six single lepton polarization
asymmetries and nine double lepton polarization asymmetries). With
this in mind a comprehensive study of polarized FB asymmetries was
done by Aliev {\sl et al.} \cite{Aliev:2004hi}. Inclusion of all these
observables would give us 33 observables with which to fix the 24
parameters. So even if some of our observables are small we should
still have sufficiently many observables to constrain the value of our
24 parameters. 

%%%%%%%%%%%%%
\par To summarize, the various polarization asymmetries
show a strong dependence on the scalar and tensorial interactions.
Also the phase of the Wilson coefficients can give substantial
deviations in the polarization asymmetries. This is of great
importance as various polarization asymmetries have different
bilinear combinations of Wilsons and hence have independent
information. Hence they can be very useful in not only estimating
the magnitude of the various Wilson coefficients but also in
providing information regarding their phases.
\vskip 2cm

%%%%%%%%%%%%%%%%%%%%%%%
%  Figures

%%%%%%%%%%%%%%%%%%%%%%%%%%%%%%%%%%%%%%%%
\FIGURE[ht]{
\epsfig{file=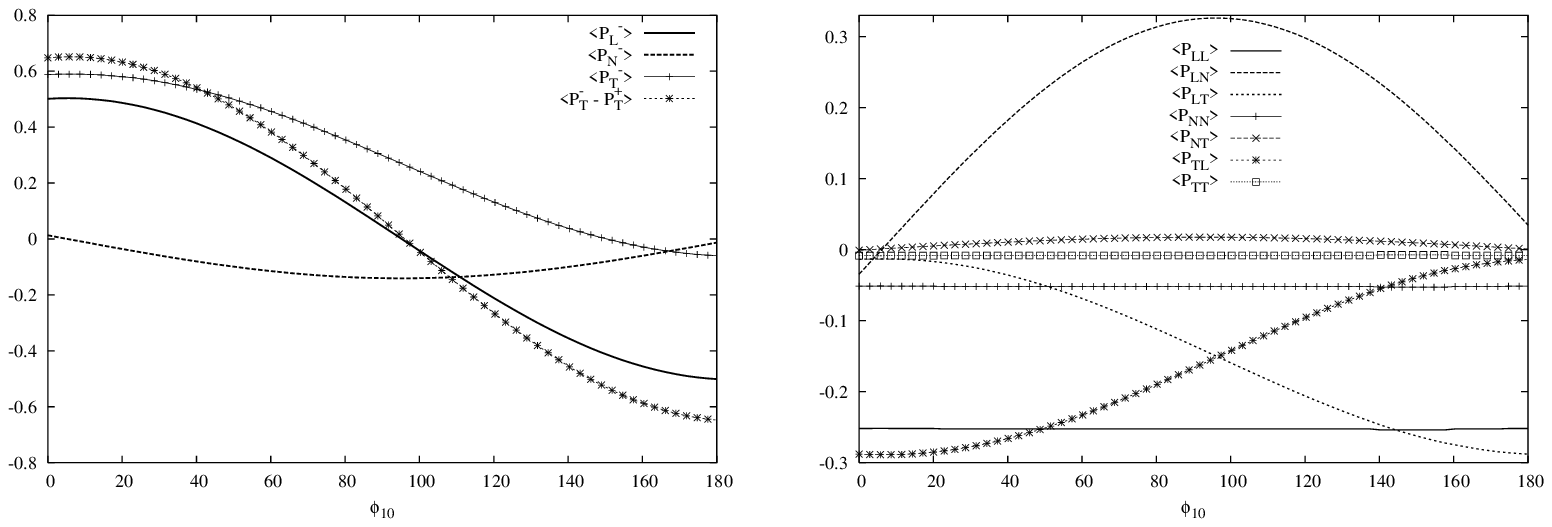,width=.92\textwidth} \caption{The
variation of the lepton polarization asymmetries as a function of
the phase (in degrees), $\phi_{10}$, of $C_{10}$ where we have
taken $|C_{10}| = 4.669$.} \label{fig:11}
}
%%%%%%%%%%%%%%%%%%%%%%%%%%%%%%%%%%%%%%%%%
%%%%%%%%%%%%%%%%%%%%%%%%%%%%%%%%%%%%%%%%
\FIGURE[ht]{
\epsfig{file=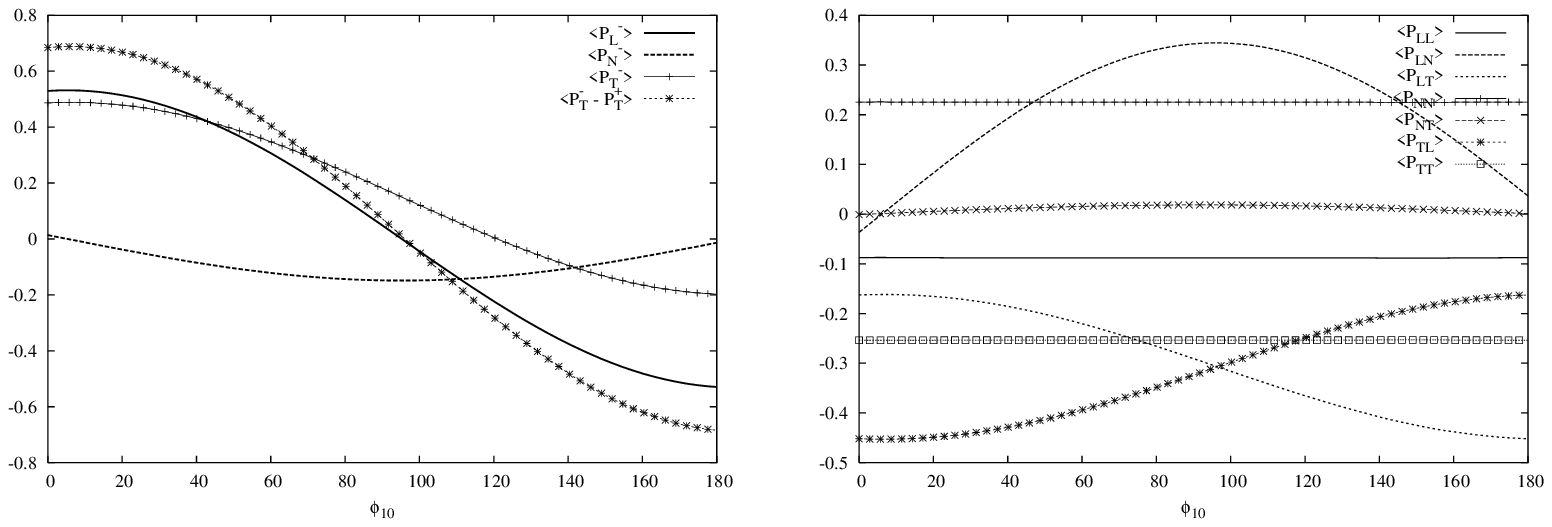,width=.92\textwidth} \caption{The
same as Figure (\ref{fig:11}) but now taking $|C_{10}| = 9$.}
\label{fig:12}
}
%%%%%%%%%%%%%%%%%%%%%%%%%%%%%%%%%%%%%%%%%
%%%%%%%%%%%%%%%%%%%%%%%%%%%%%%%%%%%%%%%%
\FIGURE[ht]{
\epsfig{file=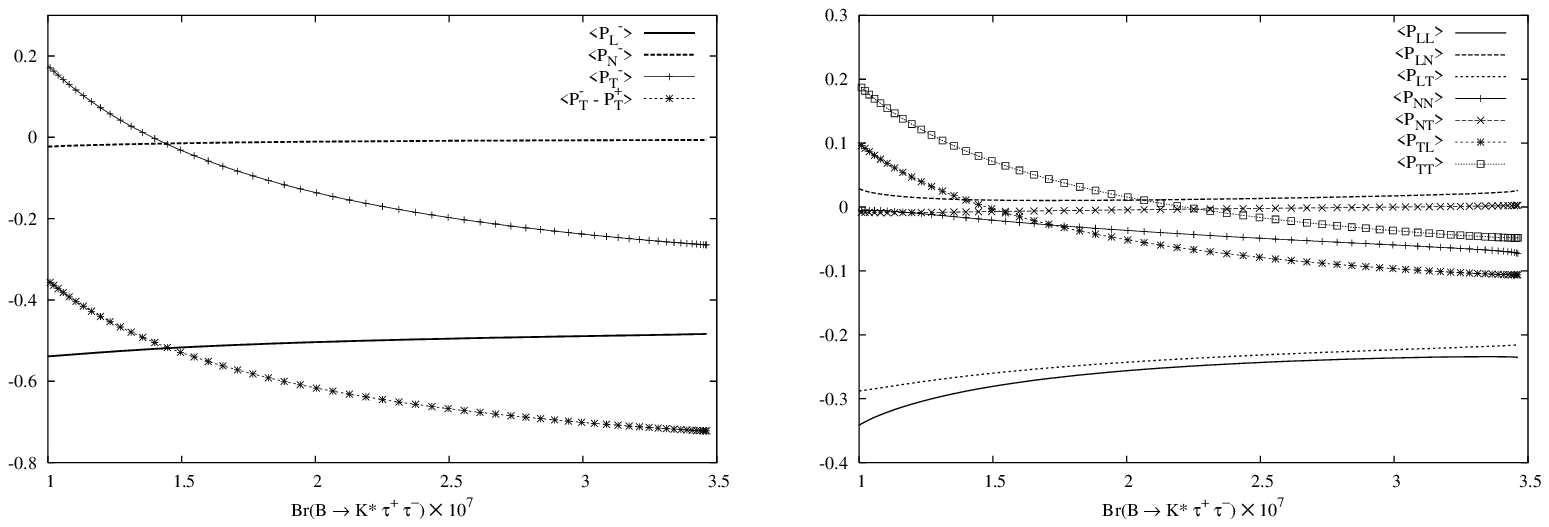,width=.92\textwidth} \caption{The
polarization asymmetries as a function of the branching ratio
varied across the phase range $0 \le \phi_{RL} \le \pi$; where the
magnitude of $C_{RL}$ is taken to be $|C_{RL}| = 4$.}
\label{fig:13}
}
%%%%%%%%%%%%%%%%%%%%%%%%%%%%%%%%%%%%%%%%%
%%%%%%%%%%%%%%%%%%%%%%%%%%%%%%%%%%%%%%%%
\FIGURE[ht]{
\epsfig{file=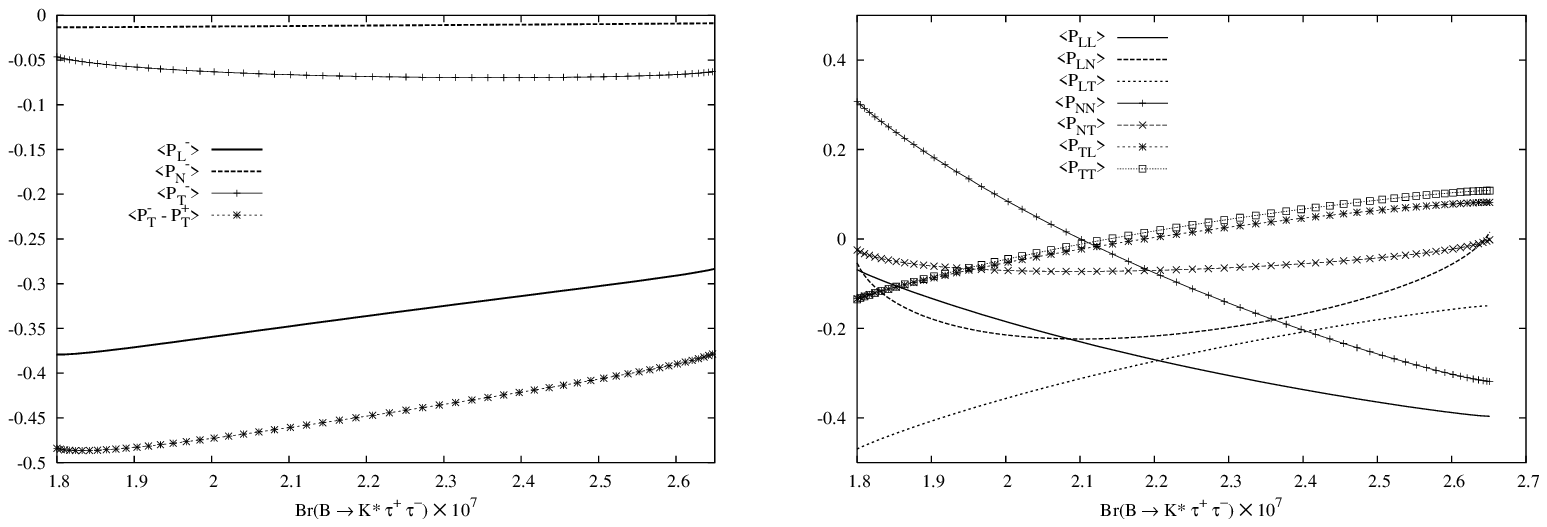,width=.92\textwidth} \caption{The
same as Figure (\ref{fig:13}) but now with $|C_{RR}|=4$ and we
have varied the phase, $\phi_{RR}$, in the range $0 \le \phi_{RR} \le
\pi$.} \label{fig:14}
}
%%%%%%%%%%%%%%%%%%%%%%%%%%%%%%%%%%%%%%%%%

%%%%%%%%%%%%%%%%%%%%%%%
%  Acknowledgements

\acknowledgments
The authors would like to thank S. Rai Choudhury for the useful
discussions during the course of the work. The authors are also
greatful to T. M. Aliev for some useful clarifications.  
\par The work of NG was supported under SERC scheme of the
Department of Science and Technology (DST), India in project no.
SP/S2/K-20/99. NG would like to thank KIAS, Korea for their
hospitality, where part of this work was also done.

%%%%%%%%%%%%%%%%%%%%%%%
%  Appendices

\appendix

\section{Input parameters}\label{inputpara}
\begin{center}
$|V_{tb} V_{ts}^*| = 0.0385 ~~,~~ \alpha = \frac{1}{129}$ \\
$G_F = 1.17 \times 10^{-5} {\rm ~~GeV^{-2}} ~~,~~ 
\Gamma_B = 4.22 \times 10^{-13} {\rm ~~GeV}$ \\ 
$m_B = 5.3 {\rm ~ GeV} ~~,~~ m_{K^*} = 0.89 {\rm ~ GeV} ~~,~~
 m_b = 4.5 {\rm ~ GeV}$
\end{center}

\section{Some Analytical Expressions}\label{appendix:a}

\subsection{Parameterization of Form Factors}\label{formfactor}

\par In our calculations we have parameterized our form
factors according to the expression;
\begin{eqnarray}
F(\hat{s}) & = & F(0) \exp \left( c_1 \hat{s} + c_2 \hat{s}^2
\right) , \label{formparam}
\end{eqnarray}
where we have used the central value of parameterization given in
Table 3 of Ali {\em et al.} \cite{Ali:1999mm}, we reproduce this
table below (Table \ref{tableform}).

% Table form factor
%\begin{table}[h]
%\begin{center}
\TABULAR{||c || c | c | c | c | c | c | c ||}
%\begin{tabular}{||c || c | c | c | c | c | c | c ||}
{\hline \hline & $A_1$ & $A_2$ & $A_0$ & $V$ & $T_1$ & $T_2$ & $T_3$ \\
\hline \hline
$F(0)$ & 0.377 & 0.282 & 0.471 & 0.457 & 0.379 & 0.379 & 0.260 \\
$c_1$ & 0.602 & 1.172 & 1.505 & 1.482 & 1.519 & 0.517 & 1.129 \\
$c_2$ & 0.258 & 0.567 & 0.710 & 1.015 & 1.030 & 0.426 & 1.128 \\
\hline }
{The parameterizing coefficients for the form factors, as
expressed in Eqn.(\ref{formparam})\label{tableform}}

%%%%%%%%%%%%%%%%%%%%%%%%%%%%

\subsection{The unpolarized cross-section}\label{delta}

The terms in the unpolarized cross-section, given in
Eqn.(\ref{sec:2:eqn:9}), are;
\begin{eqnarray}
\Delta &=&
{4 \over 3} \lambda m_B^6
\left\{ \left( 2 \mlhs + \sh \right) |A|^2
      + \left( \sh - 4 \mlhs \right) |E|^2
\right\}
+ {2 \over 3}\frac{m_B^2}{\mkhs \sh}
\Bigg[ \left(2 \mlhs + \sh \right)
\left\{
  \left(\lambda + 12 \sh \mkhs \right) |B|^2
  \right. \nonumber \\
&&\left. +  m_B^4 \lambda^2 |C|^2 \right\}
+ \left\{ 2 ( \lambda - 24 \sh \mkhs ) \mlhs
    + \sh ( \lambda + 12 \sh \mkhs )
  \right\} |F|^2   \nonumber \\
&& + m_B^4 \lambda
  \left\{ 2 \left( \lambda + 12 \mkhs \sh \right) + \sh \lambda
  \right\} |G|^2 + 2 m_B^2 (2 \mlhs + \sh ) (1 - \sh - \mkhs) \lambda
  Re(B^* C)  \nonumber \\
&& + 2 \lambda m_B^2
  \left\{
   \left(1 - \sh - \mkhs\right) \left(2 \mlhs + \sh \right) Re(G^* F)
  + 6 \sh \mlhs Re(F^* H)
  \right\}
\Bigg]  \nonumber \\
&& + \frac{1}{\mkhs} m_B^4 \lambda
\Bigg[ 8 m_B^2 \mlhs (1 - \sh - \mkhs) Re(G^* H)
+ 4 \frac{\mlh}{m_B}
 \left\{ \sh Re(H^* M) + m_B^2(1 - \sh - \mkhs)
 \right.   \nonumber \\
&& \left. \times Re(G^* M) + Re(F^* M)
 \right\} + ( \sh - 4 \mlhs ) |K|^2  + \sh |M|^2
\Bigg]  \nonumber \\
&& + \frac{16}{3} \frac{1}{\mkhs}
\left\{ \left( \sh - 4 \mlhs \right)|C_T|^2
+ 4 \left( \sh + 8 \mlhs \right) |C_{TE}|^2
\right\}
    \nonumber \\
&& \times \Bigg[ 4 \left(\lambda + 12 \sh \mkhs \right) N_1^2
  + \lambda^2 N_2^2
  - 4 m_B^2 (1 - \sh - \mkhs)
    \left( \lambda N_1 N_2 + \mkhs N_1 T_1 \right)  \nonumber \\
&&  + 16 m_B^2 \mkhs \lambda N_2 T_1
\Bigg]
+ {1024 \over 3} \frac{m_B^4}{\sh} |C_T|^2 T_1^2
  \left[ 2 \left( \lambda - 6 \mkhs \sh \right) \mlhs
          + \lambda \sh \right]  \nonumber \\
&& + {4096 \over 3} \frac{m_B^4}{\sh} |C_{TE}|^2 T_1^2
  \left[ 2 \mlhs \left(\lambda + 12 \sh \mkhs \right) + \sh \lambda
  \right] .
\label{app:b:eq1}
\end{eqnarray}

%%%%%%%%%%%%%%%%%%%%%%%
%  References

\end{document}